%% file: main.tex
\documentclass[journal]{IEEEtran}
\usepackage{cite}
\usepackage{url}

\usepackage{algorithm}
\usepackage{algpseudocode}
\usepackage{caption}

\usepackage{amssymb}
\usepackage{stfloats}
\usepackage[square, comma, sort&compress, numbers]{natbib}
\usepackage{mathrsfs}
\usepackage{subfigure}
\usepackage{bm}
\usepackage{multirow}

\usepackage[citecolor=black]{hyperref}

\ifCLASSINFOpdf
  \usepackage[pdftex]{graphicx}
\else
  \usepackage[dvips]{graphicx}
\fi

\usepackage{enumerate}
\usepackage{enumitem}

\usepackage{amsmath}
\usepackage{amsthm}

\usepackage{array}
\usepackage{url}

\usepackage{booktabs}
\usepackage{graphicx}
\usepackage{colortbl}

\usepackage{verbatim} 

\ifCLASSINFOpdf
\else
\fi
\DeclareUnicodeCharacter{2212}{-}
\begin{document}
%

\title{A Survey of Predictive Maintenance: Systems, Purposes and Approaches}

%
%
%


\author{Tianwen~Zhu,
      Yongyi~Ran,
      Xin~Zhou,
      and
      Yonggang~Wen,~\IEEEmembership{Fellow,~IEEE}

  \thanks{T. Zhu and Y. Wen are with the School of Computer Science and Engineering, Nanyang Technological University, Singapore (e-mail: tianwen001@e.ntu.edu.sg and ygwen@ntu.edu.sg). Y. Ran is with the School of Communication and Information Engineering, Chongqing University of Posts and Telecommunications, China (e-mail: ranyy@cqupt.edu.cn). X. Zhou is with the School of Information and Mechatronics Engineering, Jiangxi Science and Technology Normal University, China (e-mail: zhouxin@jxstnu.edu.cn).}
  \thanks{Corresponding author: X. Zhou.}
  }

\maketitle

\input{abstract}

\IEEEpeerreviewmaketitle

\input{intro}

\input{type}

\input{sys}

\input{obj}

\input{ml}

\input{deep}

\input{future}

\input{conclusion}

\footnotesize
\bibliography{ref}
\bibliographystyle{IEEEtran}

\end{document}

%% file: abstract.tex
\begin{abstract}

This paper highlights the importance of maintenance techniques in the coming industrial revolution, reviews the evolution of maintenance techniques, and presents a comprehensive literature review on the latest advancement of maintenance techniques, i.e., Predictive Maintenance (PdM), with emphasis on system architectures, optimization objectives, and optimization methods. In industry, any outages and unplanned downtime of machines or systems would degrade or interrupt a company's core business, potentially resulting in significant penalties and immeasurable reputation and economic loss. Existing traditional maintenance approaches, such as Reactive Maintenance (RM) and Preventive Maintenance (PM), suffer from high prevent and repair costs, inadequate or inaccurate mathematical degradation processes, and manual feature extraction. The incoming fourth industrial revolution is also demanding for a new maintenance paradigm to reduce the maintenance cost and downtime, and increase system availability and reliability. Predictive Maintenance (PdM) is envisioned the solution. In this survey, we first provide a high-level view of the PdM system architectures including PdM 4.0, Open System Architecture for Condition Based Monitoring (OSA-CBM), and cloud-enhanced PdM system. Then, we review the specific optimization objectives, which mainly comprise cost minimization, availability/reliability maximization, and multi-objective optimization. Furthermore, we present the optimization methods to achieve the aforementioned objectives, which include traditional Machine Learning (ML) based and Deep Learning (DL) based approaches. Finally, we highlight the future research directions that are critical to promote the application of DL techniques in the context of PdM.

\end{abstract}


%% file: intro.tex
\section{Introduction}
\label{intro}
Maintenance as a crucial activity in industry, with its significant impact on costs and reliability, is immensely influential to a company's ability to be competitive in low price, high quality and performance. Any unplanned downtime of machinery equipment or devices would degrade or interrupt a company's business, potentially resulting in significant penalties and unmeasurable economic and reputation loss. For instance, Amazon experienced just 49 minutes of downtime, which cost the company \$4 million in lost sales in 2013. On average, organizations lose \$138,000 per hour due to data centre downtime according to a market study by the Ponemon Institute \cite{cost_dc_outages}. It is also reported that the Operation and Maintenance (O\&M) costs for offshore wind turbines account for 20\% to 35\% of the total revenues of the generated electricity \cite{gong2014current} and maintenance expenditure in oil and gas industry costs ranging from 15\% to 70\% of total production cost \cite{bevilacqua2000analytic}. Therefore, it is critical for companies to develop a well-implemented and efficient maintenance strategy to prevent unexpected outages, improve overall reliability and reduce operating costs.


The evolution of modern techniques (e.g., Internet of things, sensing technology, artificial intelligence, etc.) reflects a transition of maintenance strategies from \emph{Reactive Maintenance (RM)} to \emph{Preventive Maintenance (PM)} to \emph{Predictive Maintenance (PdM)}. RM is only executed to restore the operating state of the equipment after a failure occurs, and thus tends to cause serious lag and results in high reactive repair costs. PM is carried out according to a planned schedule based on time or process iterations to prevent breakdown, and thus may perform unnecessary maintenance and result in high prevention costs. In order to achieve the best trade-off between the two, PdM is performed based on an online estimate of the system ``health'' and can achieve timely pre-failure interventions. PdM allows the maintenance frequency to be as low as possible to prevent unplanned RM, without incurring costs associated with doing too much PM. 

Although the concept of PdM has existed for many years, it begins to be widely accessible \cite{nguyen2015multi} recently after the emerging technologies become both seemingly capable and inexpensive enough. PdM typically involves condition monitoring, fault diagnosis, fault prognosis, and maintenance plans \cite{wang2017new}. The enabling technologies have the enhanced potential to detect, isolate, and identify the precursor and incipient faults of machinery equipment and components, monitor and predict the progression of faults, and provide decision-support or automation to develop maintenance schedules. Specifically, the emerging technologies enhance PdM in the following aspects: 
\begin{enumerate}
  \item IoT for data acquisition: IoT enables gathering a huge and increasing amount of data from multiple sensors installed on machines or components \cite{ur2018big}. 
  \item Big data techniques for data (pre-)processing: Big data techniques (e.g., data cleaning and transforming, feature extraction and fusion, etc.) have been revolutionizing intelligent maintenance by turning the big machinery data into actionable information.
  \item Advanced Deep Learning (DL) methods for fault diagnosis and prognosis: In recent years, more and more DL approaches are invented and getting matured in terms of classification and regression. The larger number of layers and neurons in a DL network allow the abstraction of complex problems and enable more accuracy of fault diagnosis and prognosis (e.g., remaining useful life prediction). At the same time, the huge amount of data is able to offset the complexity increase behind DL and improve its generalization capability. 
  \item Deep Reinforcement Learning (DRL) for decision making: The breakthrough of DRL and its variants provide a promising technique for effective control in complicated systems. DRL is able to deal with highly dynamic time-variant environments with a sophisticated state space (such as AlphaGo \cite{silver2016mastering}), which can be leveraged to provide decision support for a PdM system.
  \item Powerful hardwares for complex computing: With the rapid development of semiconductor technology, the powerful hardwares, such as graphics processing unit (GPU) and tensor processing units (TPU), can significantly expedite the evolution process and reduce the required time of DL algorithms. For example, Sun \emph{et al.} \cite{sun2019optimizing} achieve a $95$-epoch training of ImageNet/AlexNet on $512$ GPUs in $1.5$ minutes.
\end{enumerate}


PdM becomes a promising approach to decrease the downtime of machines, improve the overall reliability of systems, and reduce operating costs. However, the high complexity, automation, and flexibility of modern industrial systems bring new challenges. Specifically, three fundamental aspects should be well considered in the context of PdM:
\begin{enumerate}
  \item System architecture: With the advent of Industry 4.0, a variety of techniques have been involved in industrial systems, e.g., advanced sensing techniques, cloud computing, etc. In order to design efficient, accurate and universal maintenance systems by embracing these emerging techniques, PdM systems should: a) be compatible with various industrial standards, b) be easy to integrate with the emerging or future techniques, and c) satisfy the basic requirements of PdM, e.g., data collecting, fault diagnosis, and prognosis, etc.
  \item Optimization objective: Cost and reliability are two common purposes for PdM approaches. These different purposes are often used in insulation, and may very well be in conflict. For example, for multi-component systems, when the minimum system maintenance cost is obtained, the corresponding system reliability/availability may be too low to be acceptable \cite{wang2006reliability}. Therefore, the purposes of PdM for a specific system or component should be well jointly investigated and set.
  
  \item Optimization method: The existing approaches widely varied with the used algorithms, such as algorithms based on Artificial Neural Network (ANN), Support Vector Machine (SVM), auto-encoder, and Convolutional Neural Network (CNN), etc. Also, issues of PdM are different across industries, plants and machines. Therefore, the fault diagnosis and prognosis approaches in the context of PdM must be re-designed and tailored for specific applications.
\end{enumerate}

\subsection{Existing Surveys on Fault Diagnosis and Prognosis}
There are several published survey papers that cover different aspects of fault diagnosis and prognosis related to PdM over the past years. For example,  Zhao \emph{et al.} \cite{zhao2019deep} provide a systematic overview of DL based machine health monitoring systems, including four categories of DL architecture: auto-encoder, Deep Belief Network  (DBN), CNN and Recurrent Neural Network (RNN). Most efforts of this survey are aimed at fault identification and classification other than fault prognostics. Khan \emph{et al.} \cite{khan2018review} present a simple architecture of system health management and review the applications of auto-encoder, CNN and RNN in system health management. In addition, a series of survey papers focus on the fault diagnosis for a specific type of components or equipment, e.g., bearing \cite{zhang2019machine, hoang2019survey}, rotating machinery \cite{liu2018artificial}, building systems \cite{katipamula2005methods}, wind turbines \cite{baltazar2018review}. In \cite{zhang2019machine}, Zhang \emph{et al.} systematically summarize the existing literature employing machine learning (ML) and data mining techniques for bearing fault diagnosis. Liu \emph{et al.} \cite{liu2018artificial} provide a comprehensive review of AI algorithms in rotating machinery fault diagnosis from the perspectives of theories and industrial applications. There also exist several survey papers that focus on fault prognosis. For example, Remadna \emph{et al.} \cite{remadna2018overview} present a generic Prognostic and Health Management (PHM) architecture and the applications of DL in fault prognostics. The involved DL approaches only comprise CNN, DBN and auto-encoder. Lei \emph{et al.} \cite{lei2018machinery} deliver a review of machinery prognostics following four processes of the prognostic program, namely data acquisition, HI construction, HS division and Remaining Useful Life (RUL) prediction. The model-based and data-driven approaches for RUL prediction are also summarized. 

The aforementioned survey papers have given interesting reviews related to the field of fault diagnosis and prognosis, however, they suffer from the following limitations: 1) Most of the existing survey papers \cite{zhao2019deep, khan2018review, zhang2019machine, hoang2019survey, liu2018artificial, katipamula2005methods, baltazar2018review, remadna2018overview, lei2018machinery} only focus on reviewing the existing fault diagnosis and/or prognosis approaches, most of which are equipment specific. There is no clear way provided to select, design or implement a holistic PdM system. In contrast, our paper bridges this gap to provide a high-level view of PdM for readers. We comprehensively review the work done so far in this field from the architectural perspective and present that what kinds of modules and techniques are required in a PdM system. 2) Although PdM aims to prevent unexpected outages, improve overall reliability and reduce operating costs, there is no existing survey paper to summarize the mathematical models for the maintenance purposes of PdM. In our paper, the cost, availability/reliability and multi-objective models will be covered. 3) Due to the rapid development of DL, many new DL based approaches (e.g., generative adversarial network, transfer learning, deep reinforcement learning, etc.) have been applied in the field of PdM. Therefore, a comprehensive survey is in urgent need to cover the advancements of this field in recent years (especially from $2015$ to $2020$). 

\begin{figure*}[htbp]
\centering
\includegraphics[width=6.5in]{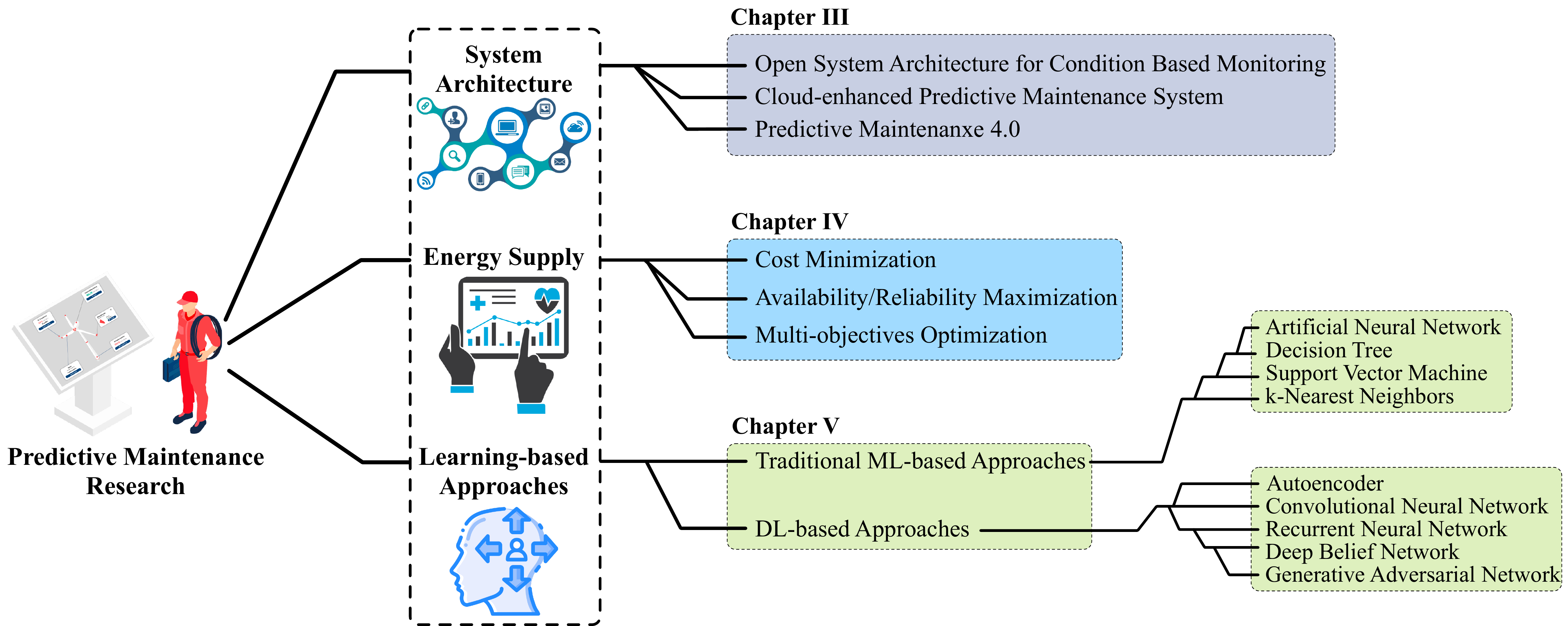}
\caption{Taxonomy of the surveyed research works.}
\label{fig:overview_survey}
\end{figure*}

\subsection{Paper Organization}

The rest of the paper is organized as below. Section II introduces the categories of maintenance, such as RM, PM and PdM. Section III presents the system architectures of PdM, which provides a high-level view of PdM. Section IV discusses the main purposes of PdM applications. In Section V, the traditional ML-based approaches are also reviewed. In Section VI, we present the DL-based PdM approaches applying in industrial equipment. Section VII focuses on future research directions. Finally, Section VIII concludes this paper. The list of abbreviations commonly appeared in this paper is given in Table \ref{tbl:abbr}.

\definecolor{mygray}{gray}{0.6}
\renewcommand\arraystretch{1.5}
\begin{table}[]
\begin{center}
\caption{List of abbreviations}
\begin{tabular}{ | p{2cm} | p{5cm}|}
     \hline
      \rowcolor{mygray} \textbf{Abbreviation} & \textbf{Description}
      \\ \hline
	  PdM
      &
	  Predictive Maintenance
      \\ \hline
      CBM
      &
      Condition-Based Maintenance
      \\ \hline
      RM
      &
      Reactive Maintenance
      \\ \hline
      PM
      &
      Preventive Maintenance
      \\ \hline
      AI
      &
      Artificial Intelligence
      \\ \hline
      DL
      &
      Deep Learning
      \\ \hline
      ML
      &
      Machine Learning
      \\ \hline
      DRL
      &
      Deep Reinforcement Learning
      \\ \hline
      IoT
      &
      Internet of things
      \\ \hline
      OSA-CBM
      &
      Open System Architecture for Condition Based Monitoring
      \\ \hline
      ANN
      &
      Artificial Neural Network
      \\ \hline
      DT
      &
      Decision Tree
      \\ \hline
      SVM
      &
      Support Vector Machine
      \\ \hline
      SVR
      &
      Support Vector Regression
      \\ \hline
      k-NN
      &
      k-Nearest Neighbors
      \\ \hline
      AE
      &
      Auto-Encoder
      \\ \hline
      CNN
      &
      Convolutional Neural Network
      \\ \hline
      RNN
      &
      Recurrent Neural Network
      \\ \hline
      DBN
      &
      Deep Belief Network
      \\ \hline
      GAN
      &
      Generative Adversarial Network
      \\ \hline
      LSTM
      &
      Long Short-Term Memory 
      \\ \hline
      GRU
      &
      Gated Recurrent Units
      \\ \hline
\end{tabular}
\label{tbl:abbr}
\end{center}
\end{table}


%% file: type.tex
\section{Evolution of Maintenance Techniques}
\label{type}
In this section, the categories of maintenance techniques will be investigated. Given the cost of downtime, a system (e.g., power system, data center) needs a well-implemented and efficient maintenance strategy to avoid unexpected outages. Nowadays, many systems still rely on spreadsheets or even pen and paper to track each piece of equipment, essentially adopting a reactive approach to upkeep. As a result, occasional downtime is expected and all too common. However, many of these outages can be prevented or minimized with the right maintenance. Maintenance strategies generally fall into one of three categories, each with its own challenges and benefits: RM, PM and PdM.

\subsection{RM}
Reactive Maintenance (RM) \cite{mobley2002introduction, swanson2001linking} is a run-to-failure maintenance management method. The maintenance action for repairing equipment is performed only when the equipment has broken down or been run to the point of failure. 

RM is appealing but few plants or companies use a true run-to-failure management philosophy. RM offers maximum utilization and in turn maximum production output of the equipment by using it to its limits. A company using run-to-failure management does not spend any money on maintenance until a machine or system fails to operate. However, the cost of repairing or replacing a component would potentially be more than the production value received by running it to failure (as shown in Fig. \ref{fig:type}). Furthermore, as components begin to vibrate, overheat and break, additional equipment damage can occur, potentially resulting in further costly repairs. In addition, a company should maintain extensive spare inventories for all critical equipment and components to react to all possible failures. The alternative is to rely on equipment vendors that can provide immediate delivery of all required spare equipment and components.

\subsection{PM}
Preventive Maintenance (PM) \cite{mobley2002introduction, wan2017manufacturing, gertsbakh2013reliability}, also referred to as planned maintenance, schedules regular maintenance activities on specific equipment to lessen the likelihood of failures. The maintenance is executed even when the machine is still working and under normal operation so that the unexpected breakdowns with the associated downtime and costs would be avoided. 

\begin{figure}[htbp]
\centering
\includegraphics[width=2.5in]{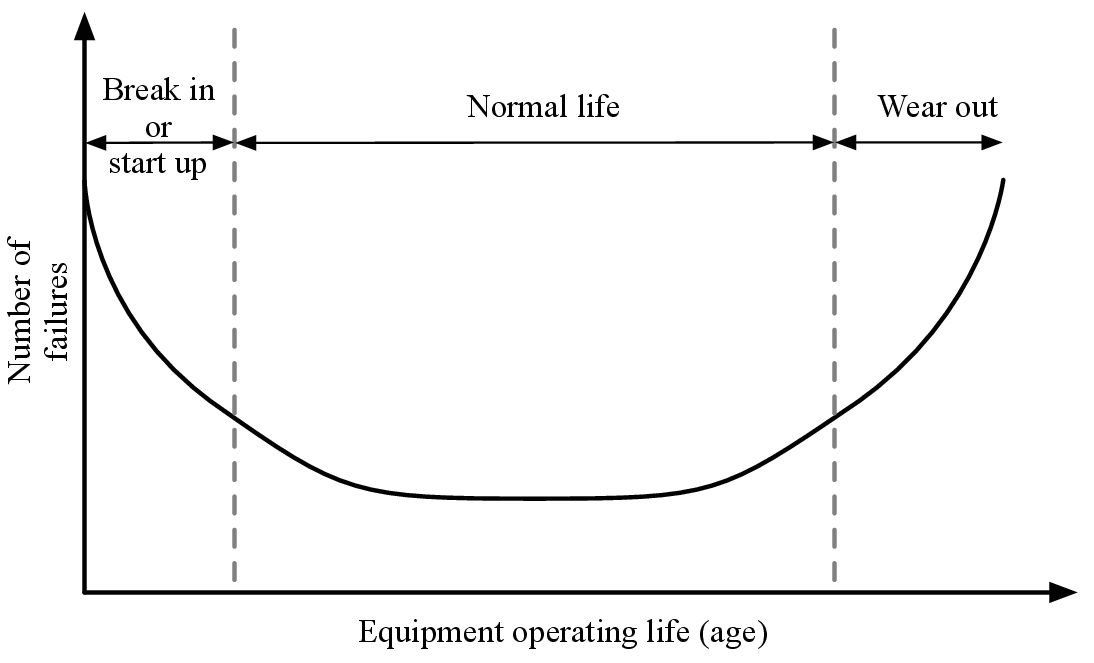}
\caption{Statistical bathtub curve of a piece of equipment \cite{mobley2002introduction, cook2018long, 8548747}.}
\label{fig:bathtub}
\end{figure}

Almost all PM management programs are time-driven \cite{mobley2002introduction, ahmad2012overview}. In other words, maintenance activities are based on elapsed time. It is assumed that the failure behavior (characteristic) of the equipment is predictable, known as bathtub curves \cite{ebeling2004introduction, mobley2002introduction, cook2018long, 8548747}, as illustrated in Fig. \ref{fig:bathtub}. The bathtub curve indicates that new equipment would experience a high probability of failure due to installation problems during the first few weeks of operation. After this break-in period, the failure rate becomes relatively low for an extended period. After this normal life period, the probability of failure increases dramatically with elapsed time. The general process of PM can be presented in two steps: 1) The first step is to statistically investigate the failure characteristics of the equipment based on the set of time series data collected. 2) The second step is to decide the optimal maintenance policies that maximize the system reliability/availability and safety performance at the lowest maintenance costs.

PM could reduce the repair costs and unplanned downtime, but might result in unnecessary repairs or catastrophic failures. Determining when a piece of equipment will enter the “wear out” phase is based on the theoretical rate of failure instead of actual stats on the condition of the specific equipment. This often results in costly and completely unnecessary maintenances taking place before there is an actual problem or after the potentially catastrophic damage has begun. Also, this will lead to much more planned downtime and require complicated inventory management. In particular, if the equipment fails before the estimated "ware out" time, it must be repaired using RM techniques. Existing analysis has shown that the maintenance cost of repairs made in a reactive mode  (i.e., after failure) is normally three times greater than that made on a scheduled basis \cite{mobley2002introduction}.

\subsection{PdM}
Predictive Maintenance (PdM), also known as condition-based maintenance (CBM) \cite{williams1994condition}, aims to predict when the equipment is likely to fail and decide which maintenance activity should be performed such that a good trade-off between maintenance frequency and cost can be achieved (as shown in Fig. \ref{fig:type}). 

The principal of PdM is to use the actual operating condition of systems and components to optimize the O\&M \cite{mobley2002introduction}. The predictive analysis is based on data collected from meters/sensors connected to machines and tools, such as vibration data, thermal images, ultrasonic data, operation availability, etc. The predictive model processes the information through predictive algorithms, discovers trends and identifies when equipment will need to be repaired or retired. Rather than running a piece of equipment or a component to failure, or replacing it when it still has useful life, PdM helps companies to optimize their strategies by conducting maintenance activities only when completely necessary. With PdM, planned and unplanned downtime, high maintenance costs, unnecessary inventory and unnecessary maintenance activities on working equipment can be decreased. However, compared with RM and PM, the cost of the condition monitoring devices (e.g., sensors) needed for PdM is often higher. Also, the PdM system is becoming more and more complex due to data collection, data analysis, and decision making.

\subsection{Summary and Comparison}
In this subsection, we summarize the differences between the three types of maintenance strategies in terms of cost, benefits, challenges, suitable and unsuitable applications.

\begin{figure}[htbp]
\centering
\includegraphics[width=3in]{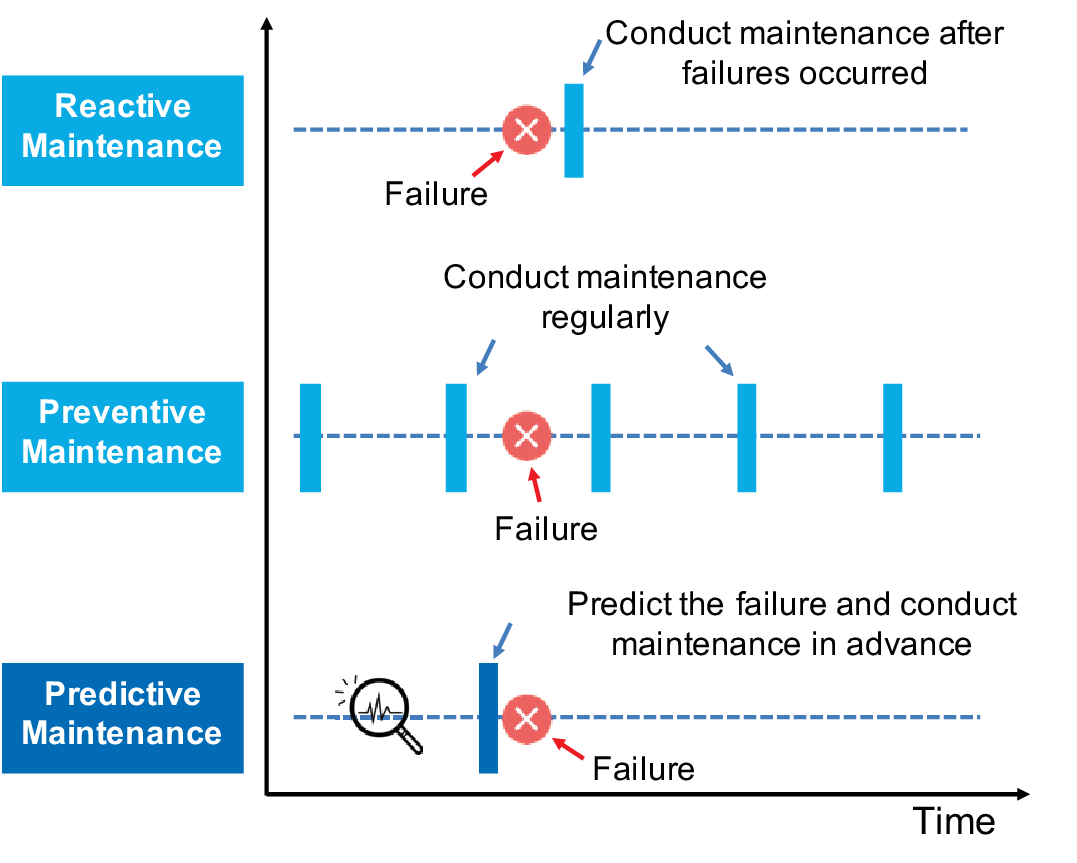}
\caption{Maintenance plans of RM, PM and PdM.}
\label{fig:plans}
\end{figure}

\begin{figure}[htbp]
\centering
\includegraphics[width=2.5in]{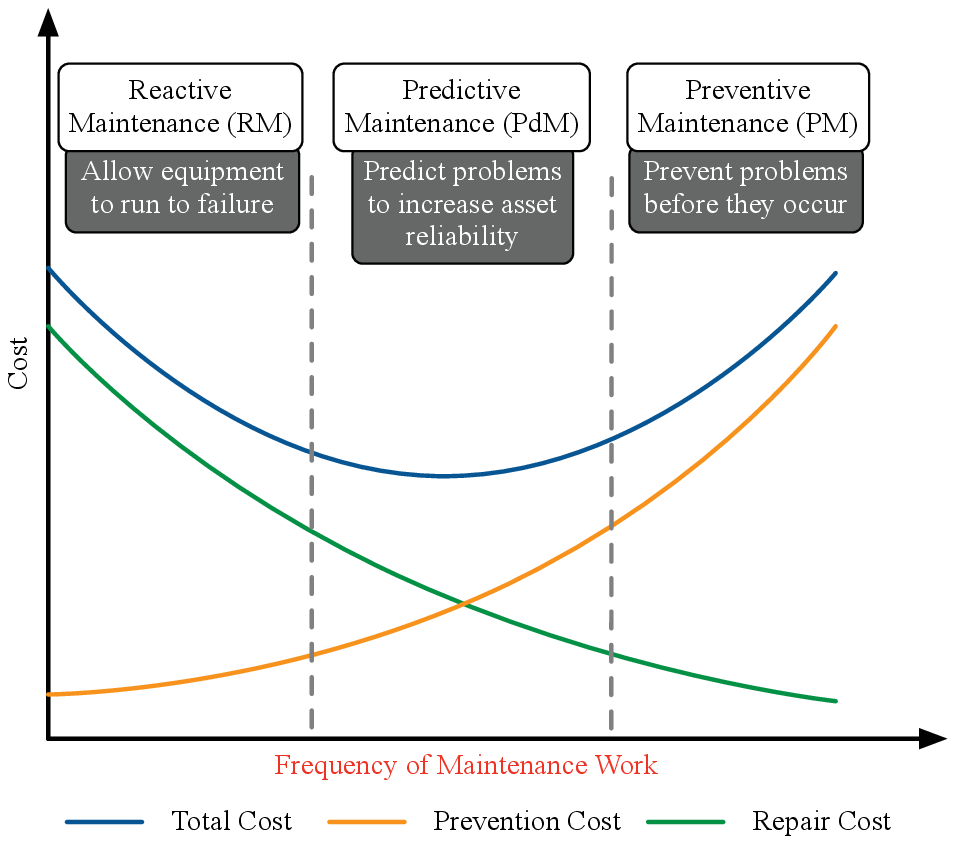}
\caption{Comparison of RM, PM and PdM on the cost and frequency of maintenance work.}
\label{fig:type}
\end{figure}

Firstly, we summarize the maintenance plans of RM, PM and PdM in Fig. \ref{fig:plans}. Then, we compare the cost \cite {mobley2002introduction} of these three maintenances in Fig. \ref{fig:type}. It can be found that RM has the lowest prevention cost due to using run-to-failure management, PM has the lowest repair cost due to well scheduled downtime while PdM can achieve the best trade-off between repair cost and prevention cost. Ideally, PdM allows the maintenance frequency to be as low as possible to prevent unplanned RM, without incurring costs associated with doing too much PM. Note that prevention cost mainly contains inspection cost, preventive replacement cost, etc., while the repair cost denotes the corrective replacement cost after failure occurred. Finally, we list the benefits, challenges, suitable and unsuitable applications for each maintenance strategy in Table \ref{tbl:type}.

\begin{table*}[]
\caption{Benefits, challenges and applications of RM, PM and PdM.}
\footnotesize
\begin{tabular}{ | p{0.4cm} | p{3.7cm} | p{3.8cm} | p{3.8cm} | p{4cm}|}
     \hline
       & Benefits & Challenges & Suitable applications & Unsuitable applications
       \\ \hline
	   RM
      & 
      \begin{itemize}[leftmargin=*]
      \item Maximum utilization and production value
      \item Lower prevention cost 
      \end{itemize}
      & 
      \begin{itemize}[leftmargin=*]
      \item Unplanned downtime
      \item High spare parts inventory cost
      \item Potential further damage for the equipment
      \item Higher repair cost 
      \end{itemize}
      &
      \begin{itemize}[leftmargin=*]
      \item Redundant, or non-critical equipment
      \item Repairing equipment with low cost after breakdown
      \end{itemize}
      &
            \begin{itemize}[leftmargin=*]
      \item Equipment failure creates a safety risk 
      \item 24/7 equipment availability is necessary
      \end{itemize}
      \\ \hline
      PM
            & 
      \begin{itemize}[leftmargin=*]
      \item Lower repair cost
      \item Less equipment malfunction and unplanned downtime 
      \end{itemize}
      & 
      \begin{itemize}[leftmargin=*]
      \item Need for inventory
      \item Increased planned downtime
      \item Maintenance on seemingly perfect equipment
      \end{itemize}
      &
      \begin{itemize}[leftmargin=*]
      \item Have a likelihood of failure that increases with time or use     
      \end{itemize}
      &
      \begin{itemize}[leftmargin=*]
      \item Have random failures that are unrelated to maintenance
      \end{itemize}
      \\ \hline
      PdM
            & 
      \begin{itemize}[leftmargin=*]
      \item A holistic view of equipment health
      \item Improved analytics options
      \item Avoid running to failure 
      \item Avoid replacing a component with useful life
      \end{itemize}
      & 
      \begin{itemize}[leftmargin=*]
      \item Increased upfront infrastructure cost and setup (e.g., sensors)
      \item More complex system
      \end{itemize}
      &
      \begin{itemize}[leftmargin=*]
      \item Have failure modes that can be cost-effectively predicted with regular monitoring
      \end{itemize}
      &
      \begin{itemize}[leftmargin=*]
      \item Do not have a failure mode that can be cost-effectively predicted
      \end{itemize}
      \\ \hline
\end{tabular}
\label{tbl:type}
\end{table*}


%
%

%% file: sys.tex
\section{System Architectures of PdM}
\label{sys}
In order to have a high-level view of PdM, here we introduce the existing reference system architectures of PdM and present what kinds of modules and techniques are needed for starting PdM.

\subsection{PdM 4.0}
PdM 4.0 \cite{cachada2018maintenance, li2016industry}, aligned with Industry 4.0 principles, paints a blueprint for intelligent PdM systems. Industry 4.0 \cite{chukwuekwe2016reliable, wang2016intelligent} is a paradigm shift in industrial processes and products propelled by intelligent information processing approaches, communication systems, future-oriented techniques, and more. The goal of industry 4.0 is to boost a machine or factory smarter. The ``smart'' does not just mean to improve production management but also reduces equipment downtime. Smart machines and factories use advanced technologies such as networking, connected devices, data analytics, and artificial intelligence to reach more efficient PdM. This shift on PdM under the context of Industry 4.0 is defined as PdM 4.0 \cite{wang2016intelligent, cachada2018maintenance}. 

PdM 4.0 employs advanced and online analysis of the collected data for the earlier detection of the occurrence of possible machine failures, and supports technicians during the maintenance interventions by providing a guided intelligent decision support. In \cite{haarman2017predictive}, maintenance strategies are classified into 4 levels that applied in modern industries:
\begin{itemize}
  \item Level 1 -- Visual inspections: this level conducts periodic physical inspections, and maintenance strategies are based solely on inspector’s expertise.
  \item Level 2 -- Instrument inspections: this level conducts periodic inspections, and maintenance strategies are based on a combination of inspector’s expertise and instrument read-outs.
  \item Level 3 -- Real-time condition monitoring: this level conducts continuous real-time monitoring of assets, and alerts are given based on pre-established rules or critical levels.
  \item Level 4 -- PdM 4.0: this level conducts continuous real-time monitoring of assets, and alerts are delivered based on predictive techniques, such as regression analysis.
\end{itemize}

\begin{figure}[htbp]
\centering
\includegraphics[width=3.2in]{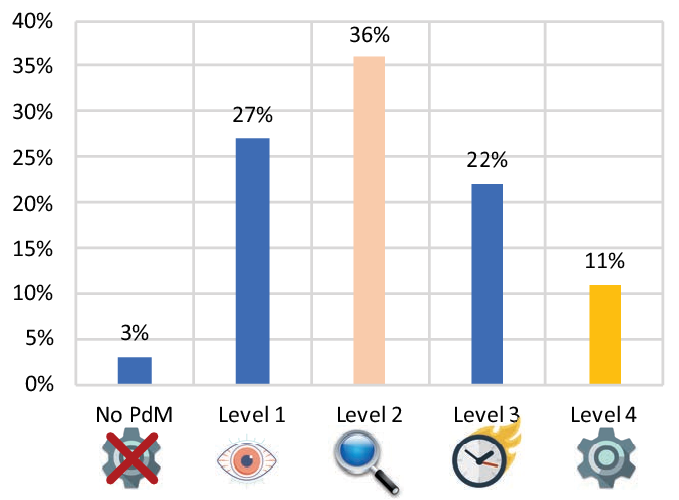}
\caption{Current predictive maintenance maturity level \cite{haarman2017predictive}. Two thirds of respondents are still below maturity level 3 for PdM. Only around 11\% have already reached level 4.}
\label{fig:pdm_percentage}
\end{figure}

In addition, as shown in Fig. \ref{fig:pdm_percentage}, \cite{haarman2017predictive} conducts a survey and finds that two thirds of respondents are still below maturity level 3. Only around 11\% have already reached level 4. These results show that PdM 4.0 has a great potential market in the near future. Therefore, here we present PdM 4.0 from a high-level and systematic perspective and point out what PdM 4.0 is and what kinds of technologies are involved in.

\begin{figure*}[htbp]
\centering
\includegraphics[width=5.5in]{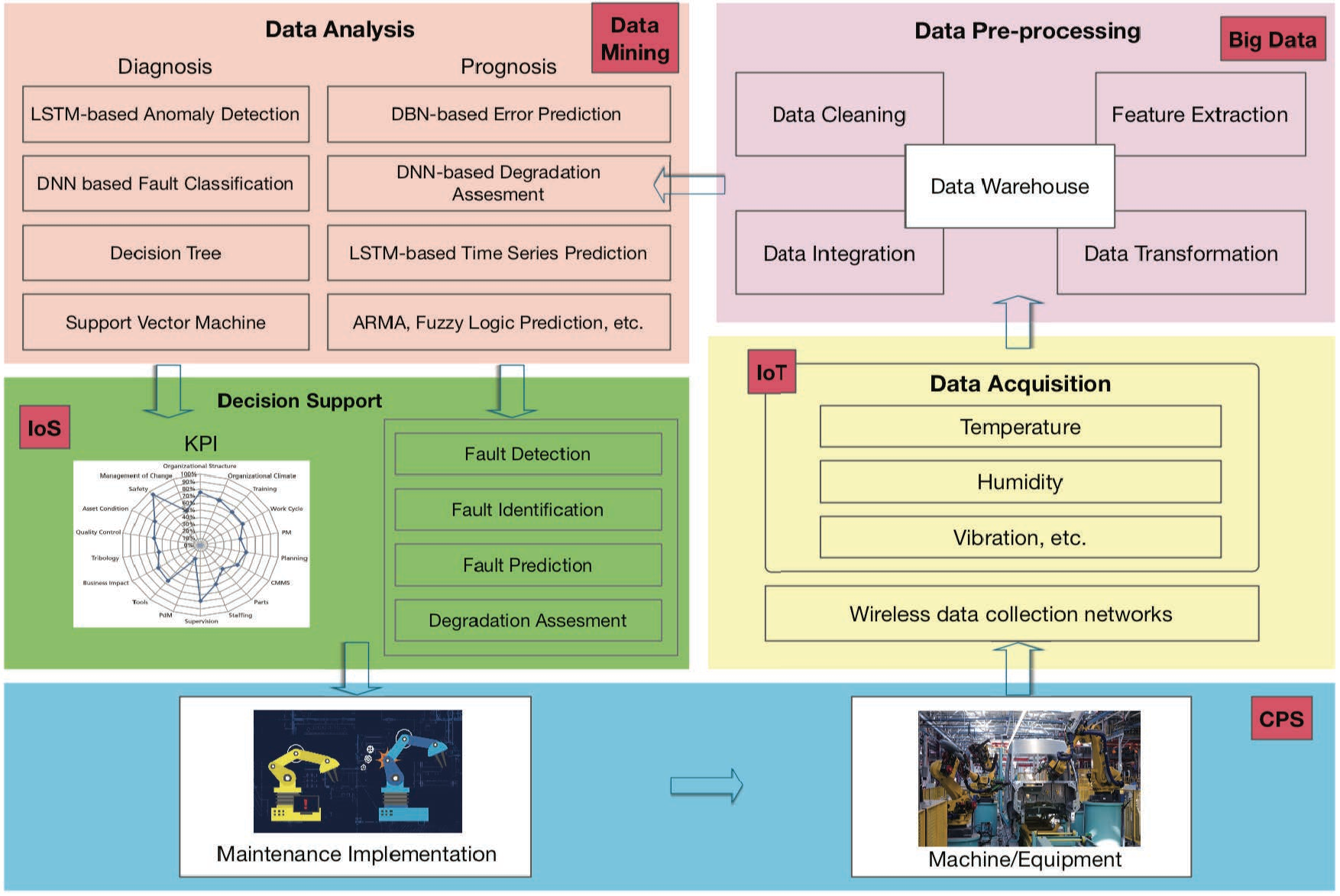}
\caption{System architecture for the intelligent and PdM 4.0. (adapted from \cite{cachada2018maintenance,li2017intelligent}).}
\label{fig:pdm40}
\end{figure*}

The system architecture for PdM 4.0 is proposed in \cite{wang2016intelligent, cachada2018maintenance}. As illustrated in Fig. \ref{fig:pdm40}, the proposed system architecture integrates the emerging advanced technologies to create a functional system that allows the implementation of intelligent PdM. The system functionality is initiated with the ``Data Acquisition'' module, where the data from several sources is collected via ``wireless sensor network'' and stored in a data warehouse. Then the data will be fed to the ``Data Pre-processing'' module, where data cleaning, data integration, data transformation and feature extraction are conducted. The output of this module will be used as the input of ``Data Analysis'' module, where advanced data analytics and machine/deep learning are used to perform the knowledge generation. The ``Decision Support'' module will visualize the result of the ``Data Analysis'' module and provide an optimized maintenance schedule. Finally, the ``Maintenance Implementation'' reacts to the physical world according to the maintenance decision and implement maintenance activities to achieve a certain purpose. More details about each module can be found in \cite{wang2016intelligent, cachada2018maintenance}.

In the reference system architecture, PdM 4.0 covers areas that include numerous technologies and related paradigms. The main elements that are closely related to PdM 4.0 comprise cyber-physical systems (CPS) \cite{lee2015cyber}, IoT \cite{civerchia2017industrial}, big data, data mining (DM), Internet of services (IoS) \cite{wang2016intelligent}.
\begin{itemize}
  \item CPS \cite{baheti2011cyber}. CPS refers to a new generation of systems with integrated computational and physical capabilities that can interact with humans via computation, communication, and control. CPS has the ability to transfer the physical world into the virtual one.
  \item IoT \cite{atzori2010internet}. IoT offers ubiquitous access to entities on the Internet by using a variety of sensing, location tracking, and monitoring devices. It enables ``objects'' to interact with each other and cooperate with their ``smart'' components to achieve common aims of PdM.
  \item Big Data \cite{zhang2017big}. Big data techniques in the context of PdM mainly involve how to store and process the large and complex data collected from sensors, e.g., filtering, data compression, data validation, feature extraction.
  \item DM \cite{li2016industry}. DM aims to analyze and discover patterns, rules, and knowledge from big data collected from multiple sources. Then, optimal decisions can be made at the right time and right place according to the result of analysis.
  \item IoS \cite{wang2016intelligent}. IoS enables service vendors to offer maintenance functions as services via the Internet. IoS consists of business models, infrastructure for services, the services themselves, and participants.
\end{itemize}

PdM 4.0 is still in its infancy. The biggest evolution of PdM 4.0 is to boost the industrial maintenance more intelligent with the help of emerging technologies. However, embracing new technologies also impose many new challenges. Here we list some of important as follows:
\begin{itemize}
  \item Fusing multi-source data: Due to the development of sensor and IoT technologies, a large amount of running and monitoring data can be collected from multiple sources in real-time, which lays the foundation for the application of data-driven AI algorithms (including traditional ML and DL algorithms). How to effectively fuse multi-source data and extract useful high-level features for fault diagnosis and prognosis is still an open issue.
  \item Promoting identification/prediction accuracy: The accuracy determines whether optimal maintenance activities can be taken in advance to effectively prevent equipment failures as well as reducing costs and downtime. Therefore, it is critical to promote the accuracy of fault identification and prediction.
  \item Optimizing maintenance scheduling: Scheduling appropriate maintenance activities subject to specific costs or availability/reliability has great significance for achieving PdM automation. There is no much research yet on how to effectively combine AI-based fault diagnosis and prognosis algorithms with maintenance scheduling algorithms to achieve an end-to-end method for intelligent and automatic PdM.
\end{itemize}



\subsection{OSA-CBM}
In this subsection, an Open System Architecture for Condition Based Monitoring (OSA-CBM) defined in ISO 13374 will be presented. 

\begin{figure}[htbp]
\centering
\includegraphics[width=2in]{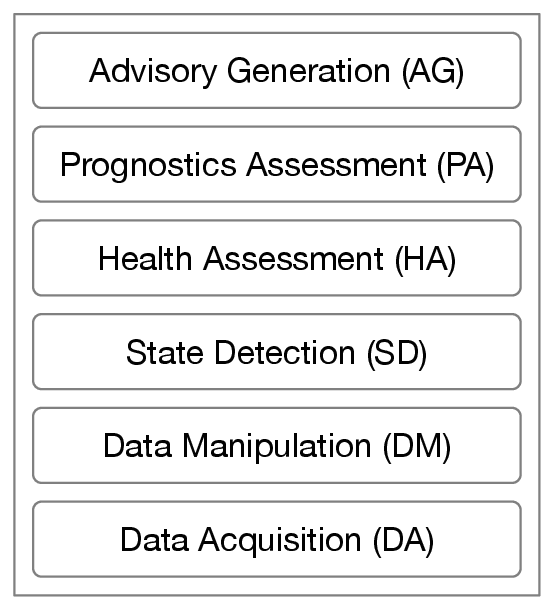}
\caption{OSA-CBM functional blocks \cite{MIMOSAOSA-CBM}.}
\label{fig:layer}
\end{figure}

OSA-CBM provides a uniform and layered framework to guide the design and implementation of a PdM system. PdM has been utilized in the industrial world since the 1990s \cite{williams1994condition}. In 2003, ISO issued a series of standards related to condition-based maintenance. Among these standards, ISO 13374 \cite{iso13374} addresses the OSA-CBM, held by Machinery Information Management Open Systems Alliance (MIMOSA \cite{MIMOSAOSA-CBM}), representing formats and methods for communicating, presenting, and displaying relevant information and data. Initially, OSA-CBM comprised seven generic layers to gain a well-constructed system \cite{lebold2002osa}, but currently considers six functional blocks \cite{MIMOSAOSA-CBM}, as shown in Fig. \ref{fig:layer}:
\begin{itemize}
  \item Data Acquisition: provides the access to the installed sensors and collects data.
  \item Data Manipulation: performs single and/or multi-channel signal transformations and applies specialized feature extraction algorithms to the collected data.
  \item State Detection: conducts condition monitoring by comparing features against expected values or operational limits and returning conditions indicators and/or alarms.
  \item Health Assessment: determines whether the system is suffering degradation by taking into account the trends in the health history, operational status and maintenance history.
  \item Prognostics Assessment: projects the current health state of the system into the future by considering an estimation of future usage profiles. 
  \item Advisory Generation: provides recommendations related to maintenance activities and modification of the system configuration, by considering operational history, current and future mission profiles and resource constraints.
\end{itemize}

In addition to OSA-CBM, there are many other standards related to PdM as illustrated in Table \ref{tbl:standards}. IEEE standards, developed under the auspices of the IEEE Standards Coordinating Committee 20 (SCC20), majorly focus on the general description of testing and diagnostic information, e.g., AI-ESTATE ( IEEE 1232) and SIMICA (IEEE 1636). ISO TC 108 (Technical Committee for Standardization of Mechanical Vibration, Shock and State Monitoring) deals with mechanical vibration and shock. The issued standards are relatively more systematic, including ISO 2041, ISO 13372, ISO 13373-1, ISO 13381-1, etc. Many other organizations and countries also have made a lot of efforts to standardizing PdM as shown in Table \ref{tbl:standards}. Based on the reviewed standards, we can find that the whole PdM framework has not yet been fully built. There exists a certain overlap in the content of standards developed by different organizations and countries. At the same time, under the background of intelligent manufacturing and Industry 4.0, the emerging technologies have not yet been involved in the standardization.

\begin{table*}
\centering
\caption{A summary of international standards related to PdM.}
\footnotesize
\begin{tabular}{lp{1.85cm}lp{12cm}} 
\toprule
  Organizations & Standards &  Year & Subject \\
  or Countries & No. & & \\
  \midrule
  \multirow{4}{*} {IEEE} & IEEE P1856 & 2017 & IEEE Draft standard framework for prognostics and health management of electronic systems \\
  & IEEE 3007.2 & 2010 & IEEE recommended practice for the maintenance of industrial and commercial power systems \\
  & IEEE 1232 & 2010 & Artificial intelligence exchange and service tie to all test environment (AI-ESTATE) \\
  & IEEE 1636 & 2009 & Software interface for maintenance information collection and analysis (SIMICA) \\
  \midrule
  \multirow{9}{*} {ISO} & ISO 13373-2 & 2016 & Condition monitoring and diagnostics of machines -- Vibration condition monitoring -- Part 2: Processing, analysis and presentation of vibration data \\
  & ISO 13381-1 & 2015 & Condition monitoring and diagnostics of machines -- Prognostics -- Part 1: General guidelines \\
  & ISO 13372 & 2012 & Condition monitoring and diagnostics of machines -- Vocabulary \\
  & ISO 2041 & 2009 & Mechanical vibration, shock and condition monitoring -- Vocabulary \\
  & ISO 13374-1 & 2003 & Condition monitoring and diagnostics of machines -- Data processing, communication and presentation -- Part 1: General guidelines  \\
  & ISO 13373-1 & 2002 & Condition monitoring and diagnostics of machines -- Vibration condition monitoring -- Part 1. General procedures \\
  \midrule
  \multirow{6}{*} {IEC} & IEC 62890 & 2016 & Life-cycle management for systems and products used in industrial-process measurement, control and automation \\
  & IEC 60706-2 & 2006 & Maintainability of equipment -- Part 2: Maintainability requirements and studies during the design and development phase \\
  & IEC 60812 & 2006 & Analysis techniques for system reliability -- Procedure for failure mode and effects analysis (FMEA) \\
  & IEC 60300-3-14 & 2004 & Dependability management -- Part 3-14: Application guide -- Maintenance and maintenance support \\
  \midrule
  \multirow{7}{*} {German} & NE 107 & 2017 & NAMUR-recommendation self-monitoring and diagnosis of field devices \\
  & VDI/VDE 2651 & 2017 & Part 1: Plant asset management (PAM) in the process industry -- Definition, model, task, benefit \\
  & VDI 2896 & 2013 & Controlling of maintenance within plant management \\
  & VDI 2895 & 2012 & Organization of maintenance -- Maintenance as a task of management \\
  & VDI 2893 & 2006 & Selection and formation of indicators for maintenance \\
  & VDI 2885 & 2003 & Standardized data for maintenance planning and determination of maintenance costs -- Data and data determination \\
  \midrule
  \multirow{7}{*} {China} & GB/T 22393 & 2015 & Condition monitoring and diagnostics of machines—General guidelines \\
  & GB/T 25742.2 & 2014 & Condition monitoring and diagnostics of machines -- Data processing, communication and presentation -- Part 2: Data processing \\
  & GB/T 25742.1 & 2010 & Condition monitoring and diagnostics of machines -- Data processing, communication and presentation -- Part 1: General guidelines \\
  & GB/T 26221 & 2010 & Condition - based maintenance system architecture \\
  & GB/T 23713.1 & 2009 & Condition monitoring and diagnostics of machines -- Prognostics -- Part 1: General guidelines \\
  \bottomrule
  \end{tabular}
  \label{tbl:standards}
\end{table*}

\subsection{Cloud-enhanced PdM System}
Cloud-enhanced PdM is inspired by the potential of cloud computing \cite{Zhang2010,7177115} and cloud manufacturing \cite{xu2012cloud}. Cloud computing enables that IT resources such as infrastructure, platform, and applications are delivered as services, while cloud manufacturing transforms manufacturing resources and capabilities into manufacturing services, and offers adaptive, secure, and on-demand manufacturing services over IoT. As an important part of cloud manufacturing, PdM is enhanced by the cloud concept to support companies or plants in deploying and managing PdM services over the Internet \cite{gao2015cloud, wang2017new, mourtzis2016cloud}.


%
%

\begin{figure}[htbp]
\centering
\includegraphics[width=3.5in]{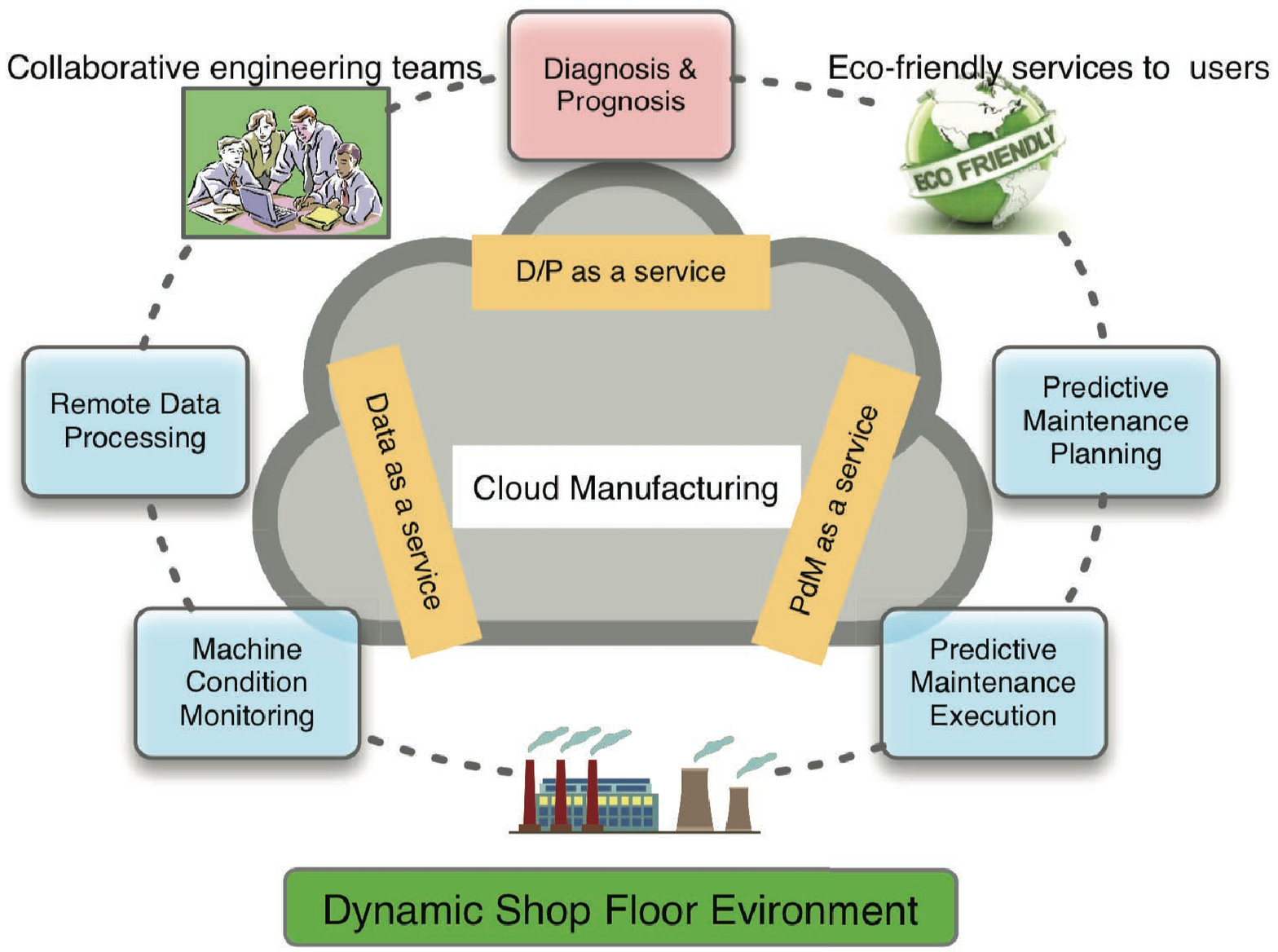}
\caption{ Architecture of cloud-enabled PdM (adapted from \cite{gao2015cloud,mourtzis2016cloud}).}
\label{fig:cloud}
\end{figure}

A generic architecture of cloud-enabled PdM is illustrated in Fig. \ref{fig:cloud}. First, machine condition monitoring collets data remotely and dynamically from the shop floor via equipped sensors. Then, remote data processing is in charge of data cleaning, data integration and feature extraction, etc. Furthermore, diagnosis and prognosis are then performed to identify and predict the potential failures respectively. The results of diagnostic and prognostic services form the basis of PdM planning, which can be remotely and dynamically executed on the shop floor. In this loop, collaborative engineering teams can provide expert knowledge in the cloud, which forms the knowledge base and can be referenced by users via the Internet. Based on this system architecture, connected equipment can deliver data-as-a-Service to the cloud-enhanced PdM. On the other hand, equipment can subscribe prognosis-as-a-service or in more general case maintenance-as-a-service. Besides cloud computing and cloud manufacturing, the supporting technologies to implement cloud-enabled PdM successfully also contain IoT, embedded system, semantic web, and machine-to-machine communication, etc. For example, Mourtzis \emph{et. al} \cite{mourtzis2018cloud} integrate IoT as well as sensor techniques in their proposed cloud-based cyber-physical system for adaptive shop-floor scheduling and condition-based maintenance. Edrington \emph{et. al} \cite{edrington2014machine} apply MTConnect \cite{MTConnect} technology to achieve data collection, analysis, and machine event notification in their web-based machine monitoring system.

From the perspective of PdM, cloud computing environment can efficiently support various smart services and solve several issues such as the memory capacity of equipment, computing power of processor, data security and data fusion from multiple sources. Therefore, such cloud-enhanced PdM paradigm possesses the following characteristics \cite{wang2017new}:
\begin{enumerate}
  \item Service-oriented. All PdM functions can be derived as cloud-based services. The users no longer need to host and maintain a large number of computing servers or related software.
  \item Accessible and robust. The pay-as-you-go maintenance services via the Internet can increase the accessibility while the modular and configurable services can increase robustness and adaptability. 
  \item Resource-aware. The monitoring data storage and analytics computation can be performed locally or remotely, it should be resource-aware to make maintenance decisions for reducing the amount of data transmission.
  \item Collaborative and distributive. Cloud computing facilitates the sharing and exchange of information among different applications/machines at different locations seamlessly and collaboratively.
\end{enumerate}

Cloud-enhanced maintenance systems provide a new paradigm of maintenance provisioning, i.e., maintenance-as-a-service. However, a variety of challenges are involved and remain to be investigated. For example, heterogeneous data storage and analysis, communication security and user privacy. With the increasing amount of data and the development of new technologies (e.g., DL, big data analytics), the cloud-enhanced systems are further evolving to meet the demand of PdM in Industry 4.0 era.

\subsection{Digital Twin Driven PdM Framework}
The digital twin can be described as an ultra-high fidelity virtual digital copy of a real factory, a machine and more, which simulates, mirrors, predicts and improves the life of its physical entity. With the development of sensor, IoT and computation technologies, digital twin-based PdM has been attracted more and more attention. 

\begin{figure*}[htbp]
\centering
\includegraphics[width=5in]{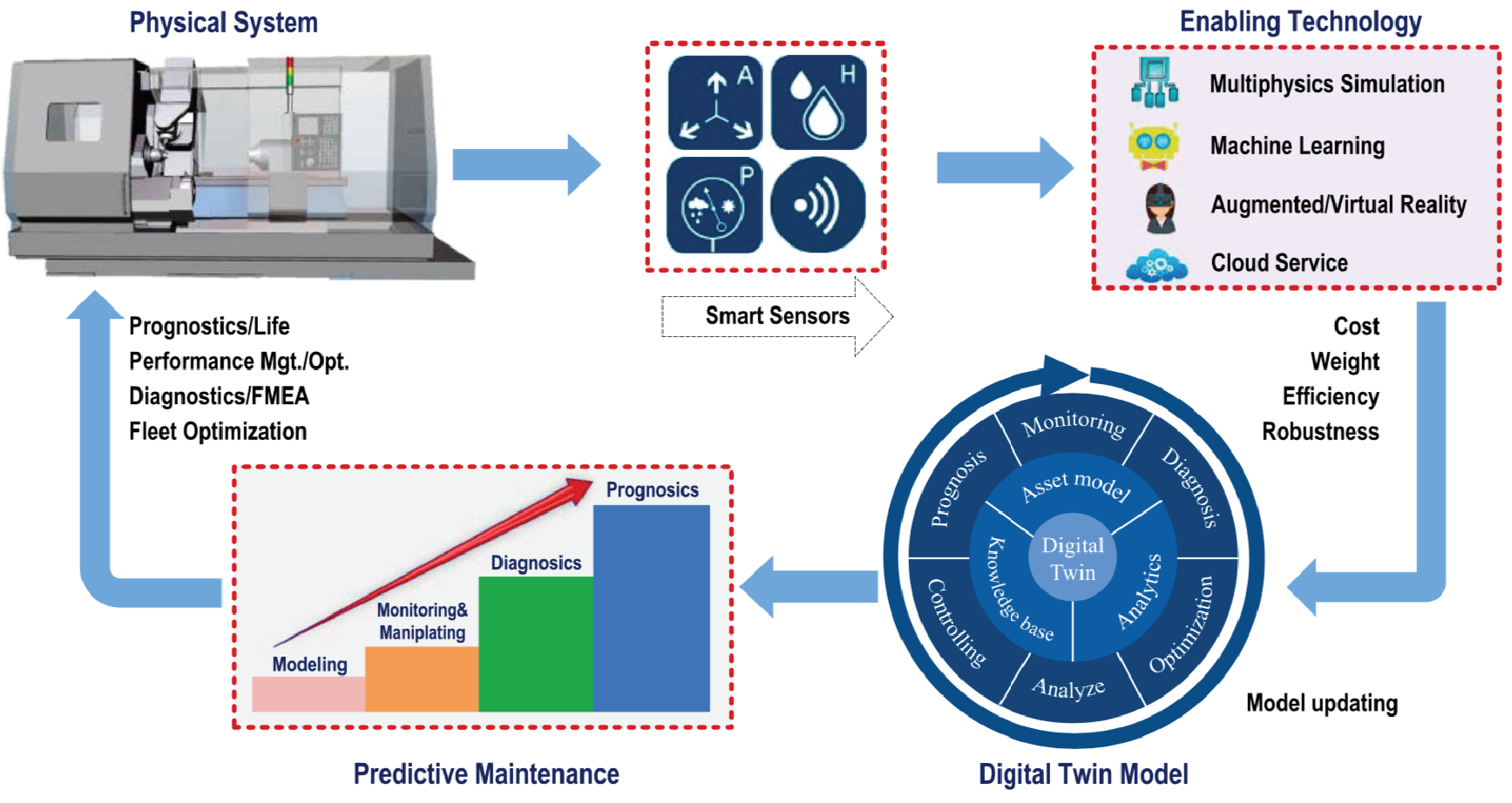}
\caption{Digital twin driven PdM framework \cite{wang2018digital}.}
\label{fig:digital_twin_pdm}
\end{figure*}

In \cite{wang2018digital}, Wang \emph{et. al}  present a ``Digital Twin'' reference model for rotating machinery fault diagnosis as shown in Fig. \ref{fig:digital_twin_pdm}. The ``Digital Twin'' reference model mainly contains three parts: (a) a physical system in Real Space, (b) a digital model in Virtual Space, and (c) the connection of data and information that ties the digital model and physical system together. On the physical side, the attributes of manufacturing system are collected from diverse sources by leveraging smart sensing techniques, ranging from the geometrical structure, material properties, process parameters, working status, to operating and environmental conditions. On the virtual side, the digital twin model is constructed by employing physics-based models and data driven analytics. The model updating technology based on parameter sensitivity analysis is used to realize its dynamic updating of the digital model according to the working status and operating conditions of the physical system. Many other efforts also have applied digital twin to assist PdM. In \cite{8598879}, the authors present a two-phase digital-twin-assisted fault diagnosis method using deep transfer learning, which realizes fault diagnosis both in the development and maintenance phases. Werner \emph{et. al} \cite{werner2019approach} develop an approach for a holistic predictive maintenance strategy by incorporating a digital twin. Other than constructing a digital twin from deterministic physics based simulations, Booyse \emph{et. al} \cite{booyse2020deep} develop digital twins from deep generative models which learn the distribution of healthy data directly from operational data at the beginning of an asset's life-cycle. 

\subsection{Miscellaneous}
Maintenance is a crucial issue to ensure production efficiency and reduce cost, thus many other PdM systems also have been proposed for different scenarios. Sipos \emph{et. al} \cite{sipos2014log} propose a log-based PdM system which utilizes state-of-the-art ML techniques to build predictive models from log data. The Senseye company \cite{Senseye} provides a system that gathers data from several sources, analyzes this data and sends a notification to a designated person when an abnormality is detected or a failure is predicted. This solution uses ML to perform condition monitoring and prognosis analysis. IBM \cite{negandhi2015ibm} provides Predictive Maintenance and Quality (PMQ) solution to help the customers monitor, analyze, and report on information that is gathered from devices and other assets and recommend maintenance activities. Many other major companies, e.g., SAP, SIEMENS and Microsoft, also have their own maintenance solution.

%% file: obj.tex
\section{Purposes of PdM}
\label{obj}
The primary purposes of PdM are to eliminate unexpected downtime, improve overall availability/reliability of systems and reduce operating costs. However, as far as we know, there is no survey paper to summarize the corresponding models for optimizing the PdM strategy so far. In this section, the literature review of optimizing PdM strategy is summarized in Table \ref{tbl:purpose} and then three categories of optimization criterions, including cost minimization, reliability or availability maximization, and multi-objective optimization, are discussed in detail. 

\begin{table*}
\centering
\caption{Literature review of optimizing PdM strategy.}
\footnotesize
\begin{center}
\begin{tabular}{llp{4cm}p{3cm}p{2cm}} 
\toprule
  Ref. & Year & Objective  &  Equipment  & Solving \\
  & & Function & & Methodologies \\
  \midrule
  Omshi \emph{et. al} \cite{omshi2020dynamic} & 2020 & Average cost  & Single-unit systems & Bayes’ theorem \\
  You \emph{et al.} \cite{you2009cost} & 2009 & Maintenance cost rate  &  Drill bit & Heuristic\\
  Grall \emph{et al.} \cite{grall2002continuous} & 2002 & Long-run s-expected maintenance cost rate  & Deteriorating systems & Heuristic \\
  He and Gu \emph{et al.} \cite{gu2017product, he2018cost, gu2017comprehensive} & 2018 & Cumulative comprehensive cost  & Cyber manufacturing systems & Heuristic \\
  Van der Weide \emph{et al.} \cite{van2010discounted} & 2010 & Discounted maintenance cost & Engineering systems with shocks & -- \\
  Louhichi \emph{et al.} \cite{louhichi2019cost} & 2019 & Total maintenance cost & Generic industrial systems  & Heuristic \\
  Feng \emph{et al.} \cite{feng2016reliability} & 2016 & Maintenance cost & Degrading Components & -- \\
  Huang \emph{et al.} \cite{huang2019degradation} & 2019 & Reliability & Dynamic environment with shocks  & Kaplan-Meier method \\
  Shen \emph{et al.} \cite{shen2018reliability} & 2018 & Reliability & Multi-component in series  & --\\
  Li \emph{et al.} \cite{li2019time} & 2019 & Reliability & Deteriorating structures & Phase-type (PH) distributions \\
  Song \emph{et al.} \cite{song2016reliability} & 2016 & Reliability & Multiple-component series systems & -- \\
  Gao \emph{et al.} \cite{gao2019reliability} & 2019 & Reliability & Degradation-shock dependence systems & -- \\
  Done \emph{et. al}\cite{dong2020reliability} & 2020 & Reliability & Parallel systems & Nelder-Mead downhill simple \\
  Yousefi \emph{et. al}\cite{yousefi2020dynamic} & 2020 & Reliability & Multi-component systems & -- \\
  Gravette \emph{et al.} \cite{gravette2015achieved} & 2015 & Availability & A US Air Force system & -- \\
  Chouikhi \emph{et al.} \cite{chouikhi2012condition} & 2012 & Availability & Continuously degrading system & Nelder-Mead method \\
  Qiu \emph{et al.} \cite{qiu2017availability} & 2017 & Steady-state availability/average long-run cost rate & Remote power feeding system & Heuristic\\
  Zhu \emph{et al.} \cite{zhu2010availability} & 2010 & Availability with cost constraints & A competing risk system & -- \\
  Compare \emph{et al.} \cite{compare2017availability} & 2017 & Availability & PHM-Equipped component & Chebyshev's inequality \\
  Tian \emph{et al.} \cite{tian2012condition} & 2012 & Cost \& reliability & Shear pump bearings & Physical programming approach \\
  Lin \emph{et al.} \cite{lin2018multi} & 2018 & Maintenance cost \& availability with reliability constraint & Aircraft fleet & Two-models-fusion \\
  Zhao \emph{et al.} \cite{zhao2018bi} & 2018 & Maintenance costs \& ship reliability & Ship & NSGA-II \\
  Xiang \emph{et al.} \cite{xiang2016multi} & 2016 & Operational cost rate \& average availability & Manufactured components & Rosenbrock method \\
  Wang \emph{et al.} \cite{wang2019vehicle} & 2019 & Total workload, total cost and demand satisfaction & Vehicle fleet & NSGA-III, SMS-EMOA, DI-MOEA \\
  Kim \emph{et al.} \cite{kim2018multi} & 2018 & Expected damage detection delay, expected maintenance delay, damage detection time-based reliability index, expected service life extension, and expected life-cycle cost & Deteriorating structure & Genetic Algorithm \\
  Rinaldi \emph{et al.} \cite{rinaldi2019multi} & 2019 & Costs, reliability, availability, costs/reliability ratio, and costs/availability ratio & Offshore wind farm & Genetic algorithms \\
  Yang \emph{et. al}\cite{yang2020theoretical} & 2020 & Risk and cost & A structural system & -- \\
  \bottomrule
  \end{tabular}
\end{center}
\label{tbl:purpose}
\end{table*}



\subsection{Cost Minimization}
The cost model varies with the applied maintenance strategy. For RM strategy, maintenance action for repairing equipment is performed only when the equipment has broken down or been run to the point of failure, thus there only exists corrective replacement cost ($C_c$). For PM strategy, sequential maintenance actions are scheduled \cite{omshi2020dynamic, you2009cost, grall2002continuous}, the involved cost items often consist of preventive replacement cost ($C_p$), inspection cost ($C_i$), unit downtime cost ($C_d$) as well as the corrective replacement cost ($C_c$). Specifically, in \cite{grall2002continuous}, Grall \emph{et al.} propose a cost model applying these cost items for continuous-time PM that aims at finding optimal preventive replacement threshold and inspection schedule based on system state. The objective cost function is to minimize the long run s-expected cost rate $EC_\infty$. Let $N_i(t)$, $N_p(t)$, and $N_c(t)$ indicate the number of inspections, preventive repairs, and corrective repairs in $[0,t]$, respectively. Let $d(t)$ denote the downtime duration in $[0,t]$. The cumulative maintenance cost can be expressed as \cite{grall2002continuous}:
\begin{equation}
  C(t)\equiv C_iN_i(t)+C_pN_p(t)+C_cN_c(t)+C_dd(t),
  \label{eq:pm_cost}
\end{equation}
and thus $EC_\infty$ is
\begin{multline} \label{eq:expected}
    EC_{\infty} = \lim_{t\rightarrow \infty} \Big[\frac{E[C(t)]}{t}\Big]\\
    \shoveleft{\qquad \ = C_i\lim_{t\rightarrow \infty} \Big[\frac{E[N_i(t)]}{t}\Big]+C_p\lim_{t\rightarrow \infty} \Big[\frac{E[N_p(t)]}{t}\Big]} \\
    \shoveleft{\quad \ + C_c\lim_{t\rightarrow \infty} \Big[\frac{E[N_c(t)]}{t}\Big]+C_d\lim_{t\rightarrow \infty} \Big[\frac{E[d(t)]}{t}\Big]}.
\end{multline}

\begin{figure}[htbp]
\centering
\includegraphics[width=3.5in]{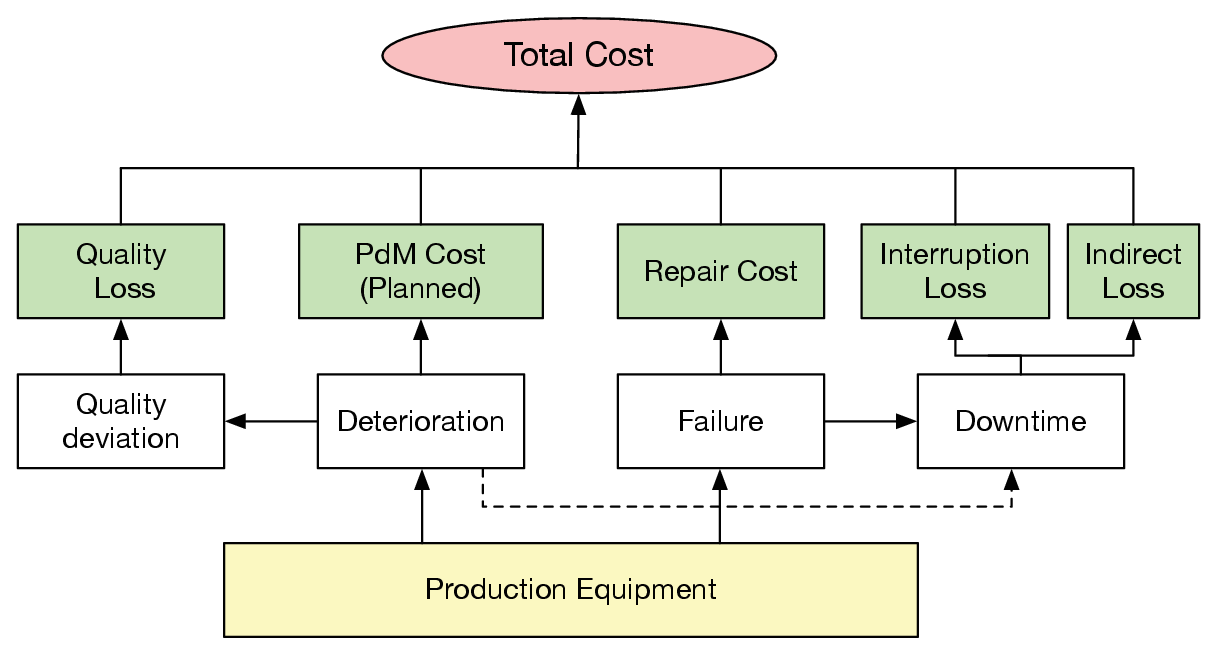}
\caption{Cost items usually used in a manufacturing system for optimizing PdM programs (adapted from \cite{gu2017product}).}
\label{fig:cost_items}
\end{figure}

For PdM strategy, maintenance actions are performed according to the failure prediction results, thus the cost model is usually associated to the RUL and depends on the specific system or equipment. In \cite{he2018cost}, He \emph{et al.} propose a comprehensive cost oriented dynamic PdM policy based on mission reliability state for a multi-state single-machine manufacturing system. As shown in Fig. \ref{fig:cost_items}, five kinds of cost items are mainly considered in this study, including corrective maintenance cost $c_1$, PdM cost $c_2$, production capacity loss cost $c_3$ caused by the maintenance actions occupying production time, indirect losses cost $c_4$ due to the production task that cannot be completed on time, and quality loss cost $c_5$. The cumulative comprehensive cost for the production task throughout the planning horizon $T$ can be given as:
\begin{equation}
  c=c_1+c_2+c_3+c_4+c_5,
  \label{eq:pdm_cost}
\end{equation}

\noindent $c_1$: Let $N$ denote the number of PdM cycles in planning horizon $T$, $\varepsilon$ represent the residual time from the last PdM activity until the end of planning horizon $T$, $\lambda_k$ indicate the failure rate of product equipment in $k$th PdM cycle, and $t_k$ represent the cumulative running time from the last implementation of PdM to the next time. Then $c_1$ can be expressed as: 
\begin{equation}
  c_1=C_c(\sum_{k=1}^N\int_0^{t_k}\lambda_k dt+\int_0^{\varepsilon}\lambda_{N+1}dt).
  \label{eq:pm_cost2}
\end{equation}

\noindent $c_2$: The cumulative PdM cost ($c_2$) throughout the planning horizon $T$ can be given as:
\begin{equation}
  c_2=\sum_{k=1}^Nc_p,
  \label{eq:pm_cost3}
\end{equation}
where $c_p$ is the expected cost for once PdM activity.

\noindent $c_3$: Equipment failures always result in production capacity loss. In \cite{he2018cost, gu2017comprehensive}, the authors gave an expression for $c_3$ according to the probability distribution of processing capacity state. Suppose the probability of processing capacity $C_x$ is $p_x$ ($x=1,2,...,M$), $C_M$ is the maximum capacity and $C_{M^{\tau'}}$ is the expected production capacity loss after a single PdM activity at the time $\tau'$. Then $c_3$ can be calculated as
\begin{equation}
  c_3=\theta(\sum_{k=1}^N\sum_{x=1,2,...,M}p_x(C_M-C_x)+\sum_{k=1}^NC_{M^{\tau'}},
  \label{eq:pm_cost4}
\end{equation}
where $\theta$ represents the expected loss cost caused by per unit production capacity reduction.

\noindent $c_4$: In practice, equipment failures may cause the production task not to be completed on time, and finally bring losses to the enterprise such as economic penalty, reputation loss and so on. Let $\sigma$ denote expected indirect economic loss, $R_T$ represent the mission reliability threshold of PdM maintenance activities in each PdM cycle for a given production task, and $R_{\varepsilon}$ indicate the the mission reliability in the residual time after the last PdM activity. Then, $c_4$ can be given as
\begin{equation}
  c_4=\sigma(\frac{\sum_{k=1}^Nt_k}{T}(1-R_T)+\frac{\varepsilon}{T}(1-R_{\varepsilon})).
  \label{eq:pm_cost5}
\end{equation}

\noindent $c_5$: As shown in Fig. \ref{fig:cost_items}, the product quality loss is caused by the deterioration of equipment, which can be represented as
\begin{equation}
  c_5=\varphi (\sum_{k=1}^Nt_k(\frac{d}{E(\rho_k)}-d)+\varepsilon(\frac{d}{E(\rho_{E+1})}-d) ),
  \label{eq:pm_cost6}
\end{equation}
where $\varphi$ indicate the economic loss caused by a single defective product, and $E(\rho_k)$ is the expected qualified rate of the equipment in the PdM cycle $k$.

More examples that apply similar cost items can be found in \cite{you2009cost, van2010discounted}. For example, in \cite{you2009cost}, You \emph{et al.} propose a cost-effective updated sequential PdM (USPM) policy to determine a real-time PM schedule for continuously monitored degrading systems. The expected maintenance cost of a system for the remaining time in the replacement cycle consists of replacement cost, imperfect PM cost and minimal CM cost. Compared with the cost model in \cite{he2018cost}, Rim \emph{et al.} \cite{louhichi2019cost} also consider ``inspect cost'' where the inspection process is performed regularly on the system to evaluate the RUL of the target system.

\subsection{Availability/Reliability Maximization}
Although cost is a good and direct criterion for judging a maintenance strategy, some parameters in the cost model are not easy to obtain, and different systems or application scenarios have different cost models. On the contrary, the uptime/downtime of the system can often be more accurately measured and easier to obtain. Therefore, availability/reliability is another practical performance metric for evaluating the effectiveness of a PdM policy.

The term ``reliability'' indicates the probability of a system or a piece of equipment to be in a functional state throughout a specified time interval \cite{letot2017updated}. Let a random variable $T_f$ represent the lifetime of the equipment, the reliability $R(t)$ is given by the following equation:
\begin{equation}
  R(t)=P(T_f>t).
  \label{eq:reliability}
\end{equation}
For $T_f$, Huang \emph{et al.} \cite{huang2019degradation} define it as the First Passage Time (FPT) that the degradation signal $\{X(t):t\geq 0\}$ exceeds a pre-specified threshold $D$, i.e., $T_f=\inf\{t\geq 0; X(t)\geq D\}$. Then, the reliability function at time $t$ can be expressed as:
\begin{equation}
  R(t)=P(T_f>t)=P(\max X(u)<D, 0\geq u\geq t).
  \label{eq:reliability2}
\end{equation}
Furthermore, the degradation signal $X(t)$ can be decomposed into two parts: a deterministic part and a stochastic part, and the reliability function is then transformed into a formula based on a Wiener process. In practice, a system usually comprises many components which are not independent. In \cite{feng2016reliability}, the authors develop a new reliability model for a complex system with dependent components subject to respective degradation processes. The dependency among components is established via environmental factors. Temperature is applied as an example application to demonstrate the proposed reliability model. Relationships between degradation and environmental temperature are studied, and then the reliability function is designed for such a system. Shen \emph{et al.} \cite{shen2018reliability} investigate the reliability of a series system with interacting components subject to continuous degradation and categorized random shocks in a recursive way and proposed a simulation method to approximate the failure time of $k$-out-of-$n$ systems. Li \emph{et al.} \cite{li2019time} propose a phase-type (PH) distribution based approach for time-dependent reliability analysis of deteriorating structures. Many other efforts that devoted to reliability analysis can be found in \cite{song2016reliability, gao2019reliability, dong2020reliability, yousefi2020dynamic}.

For ``availability'', a basic definition is given as equation (\ref{eq:availability}), which provides a probability of the system being operational. The specific availability definitions vary as what is comprised in the \emph{uptime} and \emph{downtime} \cite{gravette2015achieved}. 
\begin{equation}
  availability =\frac{uptime}{uptime+downtime}.
  \label{eq:availability}
\end{equation}

In \cite{gravette2015achieved}, the mean time between maintenances ($MTBM$) and the mean maintenance time ($M$) are employed as the uptime and downtime. Then, three generic availability models are proposed for series systems, parallel systems and series-parallel systems respectively. 

\noindent \emph{Availability for series systems ($A_a^S$)}: For a system with $n$ independent components in series, the availability ($A_a^S$) is:
\begin{equation}
  A_a^S=\Pi_{i=1}^nA_{a_i}=\Pi_{i=1}^n \frac{MTBM_i}{MTBM_i+M_i}
  \label{eq:as}
\end{equation}

\noindent \emph{Availability for parallel systems ($A_a^P$)}: For a system with $m$ independent components in parallel, the availability ($A_a^P$) is:
\begin{equation}
\begin{split}
  A_a^P & =\amalg_{j=1}^mA_{a_i}=\amalg_{j=1}^m \frac{MTBM_j}{MTBM_j+M_j} \\
  & = 1-\Pi_{j=1}^m(1-\frac{MTBM_j}{MTBM_j+M_j})
\end{split}
  \label{eq:as2}
\end{equation}

\noindent \emph{Availability for series-parallel systems ($A_a^{SP}$)}: For a system with $n$ independent subsystems connected in series and each subsystem with $m$ parallel components, the availability can be defined as:
\begin{equation}
\begin{split}
  A_a^{SP} & =\Pi_{i=1}^n[\amalg_{j=1}^mA_{a_{ij}}] \\
  & = \Pi_{i=1}^n[1-\Pi_{j=1}^m(1-\frac{MTBM_{ij}}{MTBM_{ij}+M_{ij}})]
\end{split}
  \label{eq:as3}
\end{equation}

In \cite{liao2006maintenance}, Liao \emph{et al.} develop an optimum CBM policy based on steady-state availability for a continuous degradation state considering imperfect maintenance. In \cite{chouikhi2012condition}, two kinds of system availability with/without considering the excessive degradation as unavailability are taken into account. The objective is to find an optimal inspection time for CBM and maximize the system availability. Similarly, Qiu \emph{et al.} \cite{qiu2017availability} also focus on obtaining the optimal inspection interval that maximizes the system steady-state availability and they conclude that a lower inspection interval implies an early time when the system failure can be inspected. More efforts on modeling and maximizing availability can be found in \cite{zhu2010availability, liu2016reliability, compare2017availability}.

\subsection{Multi-Objective Optimization}
Besides the aforementioned criterions, many others such as risk, safety and feasibility are commonly used in a PdM model. Usually, just one of these criterions is used as the optimization objective, e.g., minimizing maintenance cost, maximizing system reliability or minimizing equipment downtime, etc. However, such single-objective optimization approaches are often not enough to find the optimal solution that best represents the operator's preference on optimization objectives. For example, given a multi-component system, when the minimum maintenance cost is achieved, the reliability of a certain component may be too low to be acceptable. This is because that the components may be heterogeneous and diverse, the maintenance costs and degradation processes are also different. In this case, multi-objective optimization approaches are promising to achieve a better trade-off among different optimization objectives \cite{tian2012condition}. 

A general multi-objective optimization problem is to find optimum decision variables that minimize or maximize a set of different objectives. A generic mathematical formulation for multi-objective optimization problem can be given as:
\begin{multline} \label{eq:multi}
    \min f(x) =\{f_1(x), f_2(x),..., f_k(x)\} \\
    \shoveleft{s.t. \ x\in \mathcal{X}}, \qquad \qquad \qquad \qquad \qquad \qquad \qquad \qquad 
\vspace{-0.05in}
\end{multline}
where $x$ denotes the vector of decision variables, $\mathcal{X}$ is the feasible space of $x$, $k$ means the total number of objectives and $f_i(x)$ indicates the $i$-th objective function. One commonly used variant for the multi-objective optimization problem is the Weight-sum format \cite{tian2012condition}: $f(x)=\sum_{i=1}^k\omega_if_i(x)$, where $w_i$ is the weight of objective $i$ and $\sum_{i=1}^kw_i=1$, $w_i\geq0, i=1,..., k$. 

Due to the conflicting feature among different objectives, it is typically impossible to obtain the optimal values for all the objectives simultaneously. An alternative way is to find a solution that can achieve a good trade-off among the various objectives. Tian \emph{et al.} \cite{tian2012condition} formulate a multi-objective CBM optimization problem to find an optimal risk threshold value. The maintenance cost and reliability models are calculated based on proportional hazards model and a control limit CBM replacement policy, and physical programming is employed to solve the optimization problem. In \cite{lin2018multi}, Lin \emph{et al.} concurrently optimize the fleet maintenance cost and fleet availability by taking proper maintenance actions for CBM. At the same time, the failure probability calculated based on the probability–-damage–-tolerance (PDT) method is employed as a constraint. In \cite{zhao2018bi}, Zhao \emph{et al.} construct a bi-objective model under the CBM strategy to minimize the maintenance costs and maximize the ship reliability. A non-dominated sorting genetic algorithm II (NSGA-II) is employed to solve the proposed bi-objective model. Xiang \emph{et al.} \cite{xiang2016multi} considered the substantial heterogeneity in populations of manufactured components, and developed a multi-objective model to determine an optimal joint burn-in and CBM policy that minimizes the total operational cost and maximizes the average availability. More efforts on multi-objective optimization for PdM strategy can be found in \cite{lin2018multi, wang2019vehicle, kim2018multi, rinaldi2019multi, yang2020theoretical}.

%% file: ml.tex
\section{Traditional Machine Leaning based Approaches}
\label{ml}
With the development of big data techniques (e.g., sensors, IoT) and the ever-increasing size of big data, data-driven PdM is becoming more and more attractive. To extract useful knowledge and make appropriate decisions from big data, machine learning (ML) techniques have been regarded as a powerful solution. Before going ``deep'', a variety of ``shallow'' ML algorithms are developed for the context of PdM, e.g., Artificial Neural Network (ANN) \cite{samanta2003artificial, teng2016prognosis, elforjani2017prognosis, karmacharya2017fault, sharma2019bearing, jamil2015fault, chine2016novel, chine2017ann}, decision tree (DT) \cite{abdallah2018fault, li2018improved, benkercha2018fault, kou2019integrating, patil2019fault}, Support Vector Machine (SVM) \cite{santos2015svm, soualhi2015bearing, sun2016novel,  han2019least, zhu2018fault}, $k$-Nearest Neighbors ($k$-NN) \cite{appana2017reliable, baraldi2016hierarchical, uddin2016distance, tian2016motor, xiong2016information, madeti2018modeling}, particle filter \cite{orchard2009particle, daroogheh2018dual}, principal component analysis \cite{sun2007evolving, deng2018nonlinear}, adaptive resonance theory \cite{lei2013planetary, liu2017improving}, self-organizing maps \cite{zhong2005fault, rai2018intelligent}, etc. In this section, a subset of well-developed ML algorithms are reviewed and briefly summarized, with a complete list of references.

\subsection{Artificial Neural Network (ANN)}
Artificial Neural Network (ANN) is an information processing paradigm that attempts to achieve a neurological related performance, such as learning from experience, making generalizations from similar situations and judging states. In the past 3 decades, ANNs have gained much importance in fault diagnosis and prognosis. For example, machinery systems and components \cite{samanta2003artificial, teng2016prognosis, elforjani2017prognosis, karmacharya2017fault, sharma2019bearing}, and power systems \cite{jamil2015fault, chine2016novel, chine2017ann}. 

A variety of factors may affect the performance of a designed ANN for fault diagnosis and prognosis. The network architecture (e.g., the number of neurons in hidden layers, network connections, and activation functions) plays a very important role in the performance of an ANN and usually depends on the problem at hand \cite{rao2012failure}. For example, Samanta \emph{et al.} \cite{samanta2003artificial} present ANN-based fault diagnosis for rolling element bearings using time-domain features. The structure of the designed ANN giving the best results has 4-5 nodes in the input layer, 16 neurons in the first hidden layer, 10 neurons in the second hidden layer and two nodes in the output layer. Five vibration signals from sensors are used to extract five time-domain features (root mean square, variance, skewness, kurtosis and normalized sixth central moment) as the input of the designed ANN. The experimental results show that the accuracy can reach 62.5\%-100\% with different number of input signals or features. To achieve RUL prediction, Teng \emph{et al.} \cite{teng2016prognosis} and Elforjani \emph{et al.} \cite{elforjani2017prognosis} use ANN to train data-driven models and to predict the RUL of bearings. In \cite{elforjani2017prognosis}, ANN model shows good performance with a structure that has $3$ hidden layers with (7-3-7) neurons, one output layer for RUL estimation and an input layer. In addition, ANNs as ``shallow'' learning usually cannot effectively extract the informative features hidden in the raw sensors data, and thus require additional feature extraction (hand-crafted features) and/or feature selection in the learning process. Various works have reported the use of time domain \cite{samanta2003artificial, sreejith2008fault}, frequency domain \cite{sharma2019bearing} and time-frequency domain \cite{srinivasan2005artificial, karmacharya2017fault} methods to extract features.

\subsection{Decision Tree (DT)}
Decision Tree (DT) is a non-parametric supervised method used for classification and regression. DT aims to predict the value (i.e., label or class) of a target variable by learning simple decision rules inferred from the data features. A DT generally consists of a number of branches, one root, a number of interval nodes and leaves. Each path from the root node through the internal nodes to a leaf node represents a classification with the different conditions of the systems or components. Each leaf node represents a class label for classification or a response for regression. To build a DT model, one should identify the most important input variables/features, and then split instances at the root node and at subsequent internal nodes into two or more categories based on the status of such variables. C4.5 algorithm \cite{quinlan1996improved} is one of the widely used algorithms to generate decision tree.

DT-based ML techniques have been frequently utilized for PdM. First, due to the nature of DT, many efforts are devoted to identifying or classifying the state of the real-world system \cite{abdallah2018fault, li2018improved, benkercha2018fault, kou2019integrating, patil2019fault}. For example, Benkercha \emph{et al.} \cite{benkercha2018fault} propose a new approach based on DT algorithm to detect and diagnose the faults in grid connected photovoltaic system (GCPVS). The used attributes include temperature ambient, irradiation and power ratio, and the class labels contain free fault, string fault, short circuit fault or line-line fault. Experimental results indicate that the diagnosis accuracy can reach 99.80\%. Then, DT is also used to develop fault prognosis (e.g., RUL prediction) models \cite{8326010, zheng2019novel, bakir2019prediction}. For example, Bakir \emph{et al.} \cite{bakir2019prediction} apply regression tree to develop the RUL prediction model for multiple components. The results show that regression tree is simple and able to perform well even for the small dataset. Furthermore, a set of DTs can be trained and assembled to a Random Forest (RF). According to the recent literature on fault diagnosis and prognosis \cite{yan2017predictive, shi2019fault, chen2018random, wu2017comparative, patil2018remaining}, RF-based approaches are widely employed due to its low computational cost with large data and stable results.

\subsection{Support Vector Machine (SVM)}
A Support Vector Machine (SVM) is a supervised ML technique that is useful when the underlying process of the real-world system is unknown, or the mathematical relation is too expensive to be obtained due to the increased influence by a number of interdependent factors \cite{vapnik2013nature}. Typically, in the case of classification task, the samples are assumed to have two classes namely positive class and negative class, an SVM training algorithm builds a model that assigns new examples to one category or the other, making it a non-probabilistic binary linear classifier. 

Due to the high classification accuracy, even for nonlinear problems, SVM has been successfully applied to a number of applications ranging from face detection, verification and recognition to machine fault diagnosis \cite{santos2015svm, soualhi2015bearing, sun2016novel,  han2019least, zhu2018fault}. For example, Soualhi \emph{et al.} \cite{soualhi2015bearing} apply the Hilbert-Huang transform (HHT) to extract health indicator from vibration signals and utilize SVM to achieve fault classification of bearings. Support Vector Regression (SVR) is the most commonly used approaches for fault prognosis \cite{8186223, khelif2016direct, zhao2018novel, du2018battery, sui2019prediction}. In \cite{8186223}, a state-space model is built to represent battery aging dynamics using SVR. The experimental results show that the SVR-based model ensures much better performance with considerably less estimation error compared with an ANN-based model. Khelif \emph{et al.} \cite{khelif2016direct} provide a direct approach for RUL estimation determined from experiences using an SVR-RUL model. From each experience, a set of features associated with their RUL is extracted. Then, the overall set of features is fed into an SVR which aims to model the relationship between the features and the RUL. This method avoids to estimating degradation states or a failure threshold.

\subsection{k-Nearest Neighbors (k-NN)}
The $k$-nearest neighbors ($k$-NN) algorithm is a low-complexity unsupervised method that can be used for classification. The first step in k-NN classification is to determine the value of $k$, then it is necessary to compute distances (e.g. Euclidean distance) between a test instance and all training instances as a measure of similarity. The $k$-NN algorithm finds $k$ training instances that yield the minimum distances and finally assigns the test instance to the class most common among its $k$-nearest neighbors. The choice of $k$ is usually data-driven (often decided though cross-validation). Larger values of $k$ reduce the effect of noise on the classification, but lead the decision boundaries between classes less distinct. 

$k$-NN as one of the simplest approaches for classification has been widely used in the context of PdM. For example, in \cite{susto2015machine}, Susto \emph{et al.} apply $k$-NN as a classifier in their proposed Multiple Classifier (MC) PdM methodology to deal with the ``integral type faults'' problem. To obtain a better classification performance, some enhanced $k$-NN methods are proposed \cite{appana2017reliable, baraldi2016hierarchical, uddin2016distance, tian2016motor, xiong2016information, madeti2018modeling}. In \cite{appana2017reliable}, Appana \emph{et al.} present a density-weighted distance similarity metric, which considers the relative densities of samples in addition to the distances between samples to improve the classification accuracy of standard k-NN. Also, $k$-NN can be applied for RUL prediction and early fault warning. Liu \emph{et al.} \cite{liu2017remaining} propose a VKOPP model based on optimally pruned extreme learning machine (OPELM) and Volterra series to estimate the RUL of the insulated gate bipolar transistor (IGBT). This model uses a combination of $k$-NN and least squares estimation (LSE) method to calculate the output weights of OPELM and predict the RUL of the IGBT. In \cite{chen2018evidential}, Chen \emph{et al.} develop a so-called CMEW-EKNN method based on the evidential $k$-NN (EKNN) rule in the framework of evidence theory to achieve a condition monitoring and early warning in power plants.

\subsection{Summary} 
Different traditional ML methods have different characteristics and applicable scenarios. TABLE \ref{tbl:ml} briefly summarizes the learning strategies, advantages, limitations and typical applications of the commonly used traditional ML approaches for PdM.

\begin{table*}[]
\begin{center}
\caption{Advantages \& Limitations of Traditional ML based Approaches.}
     \begin{tabular}{ | p{1.5cm} | p{3.5cm} | p{3.5cm} | p{3.5cm} | p{3.5cm}|}
     \hline
      Approaches & Learning Strategies & Advantages & Limitations & Typical Applications for PdM
       \\ \hline
	  ANN
      & 
      \begin{itemize}[leftmargin=*]
      \item Iteratively update the weight parameters to minimize training loss by using the gradient descent algorithm
      \end{itemize}
      & 
      \begin{itemize}[leftmargin=*]
      \item High classification and prediction accuracy
      \item Good approximation of nonlinear function
      \end{itemize}
      & 
      \begin{itemize}[leftmargin=*]
      \item Many weight parameters need to be trained
      \item May require greater computational resources 
      \item Easy to over-fit
      \item No physical meaning
      \item No standard to decide network structure
      \end{itemize}
      &
      \begin{itemize}[leftmargin=*]
	  \item Fault diagnosis of bearings \cite{samanta2003artificial, benkercha2018fault}
	  \item Predicting RUL of bearings \cite{teng2016prognosis, elforjani2017prognosis}
	  \end{itemize}
      \\ \hline
      DT
      & 
      \begin{itemize}[leftmargin=*]
      \item Recursively split the training data and assigning a class label to leaf by the most frequent class
      \item Prune a subtree with a leaf or a branch if lower training error obtained
      \end{itemize}
      & 
      \begin{itemize}[leftmargin=*]
      \item Easy to understand
      \item Non-parametric
      \end{itemize}
      & 
      \begin{itemize}[leftmargin=*]
      \item Easy to over-fit
      \item Poor prediction accuracy
      \item Unstable
      \end{itemize}
      &
      \begin{itemize}[leftmargin=*]
      \item Fault diagnosis in GCPVS \cite{benkercha2018fault}, rail vehicles \cite{kou2019integrating}, refrigerant flow system \cite{li2018improved} and anti-friction bearing \cite{patil2019fault}, etc.
      \item Fault prognosis for turbofan engine \cite{8326010}, lithium-ion battery \cite{zheng2019novel} and mech equipment \cite{bakir2019prediction}, etc.
	  \end{itemize}
      \\ \hline
      SVM
      & 
      \begin{itemize}[leftmargin=*]
      \item Identify the optimal hyper-plane
      \item Derive classification rules from association patterns
      \end{itemize}
      & 
      \begin{itemize}[leftmargin=*]
      \item Work well with even unstructured and semi structured data
      \item Can deal with high-dimensional features
      \item The risk of over-fitting is less
      \end{itemize}
      & 
      \begin{itemize}[leftmargin=*]
      \item No probabilistic explanation for the classification
      \item No standard for choosing the kernel function
      \item Low efficiency for big data
      \end{itemize}
      &
	  \begin{itemize}[leftmargin=*]
      \item Fault diagnosis for chiller \cite{han2019least}, rotation machinery \cite{zhu2018fault}, bearings \cite{soualhi2015bearing} and wind turbines \cite{santos2015svm}, etc.
      \item RUL prediction for Lithium-Ion batteries \cite{8186223, zhao2018novel, du2018battery} and bearing \cite{sui2019prediction}, etc.
      \end{itemize}
      \\ \hline
      KNN
      & 
      \begin{itemize}[leftmargin=*]
      \item A test sample is given as the class of majority of its nearest neighbours
      \end{itemize}
      & 
      \begin{itemize}[leftmargin=*]
      \item Few parameters to tune
      \item Very easy to implement
      \item No training step
      \end{itemize}
      & 
      \begin{itemize}[leftmargin=*]
      \item $K$ must be determined in advance
      \item Sensitiveness to unbalanced datasets and noisy/irrelevant attributes
      \item Curse of dimensionality
      \end{itemize}
      &
      \begin{itemize}[leftmargin=*]
      \item Fault diagnosis \cite{susto2015machine, appana2017reliable, baraldi2016hierarchical, uddin2016distance, tian2016motor, xiong2016information, madeti2018modeling}
      \item RUL prediction \cite{liu2017remaining} and early fault warning \cite{chen2018evidential}
      \end{itemize}
      \\ \hline
      \end{tabular}
      \label{tbl:ml}
\end{center}
\end{table*}

%% file: deep.tex
\section{Deep Learning Based Approaches}
\label{deep}
Recently, deep learning has shown superior ability in feature learning, fault classification and fault prediction with multilayer nonlinear transformations. Auto-Encoder (AE), Convolutional Neural Network (CNN), Deep Belief Network (DBN), and other deep learning models are widely applied in the field of PdM.

\subsection{Auto-encoder (AE)}
\begin{figure}[htbp]
\centering
\includegraphics[width=2.2in]{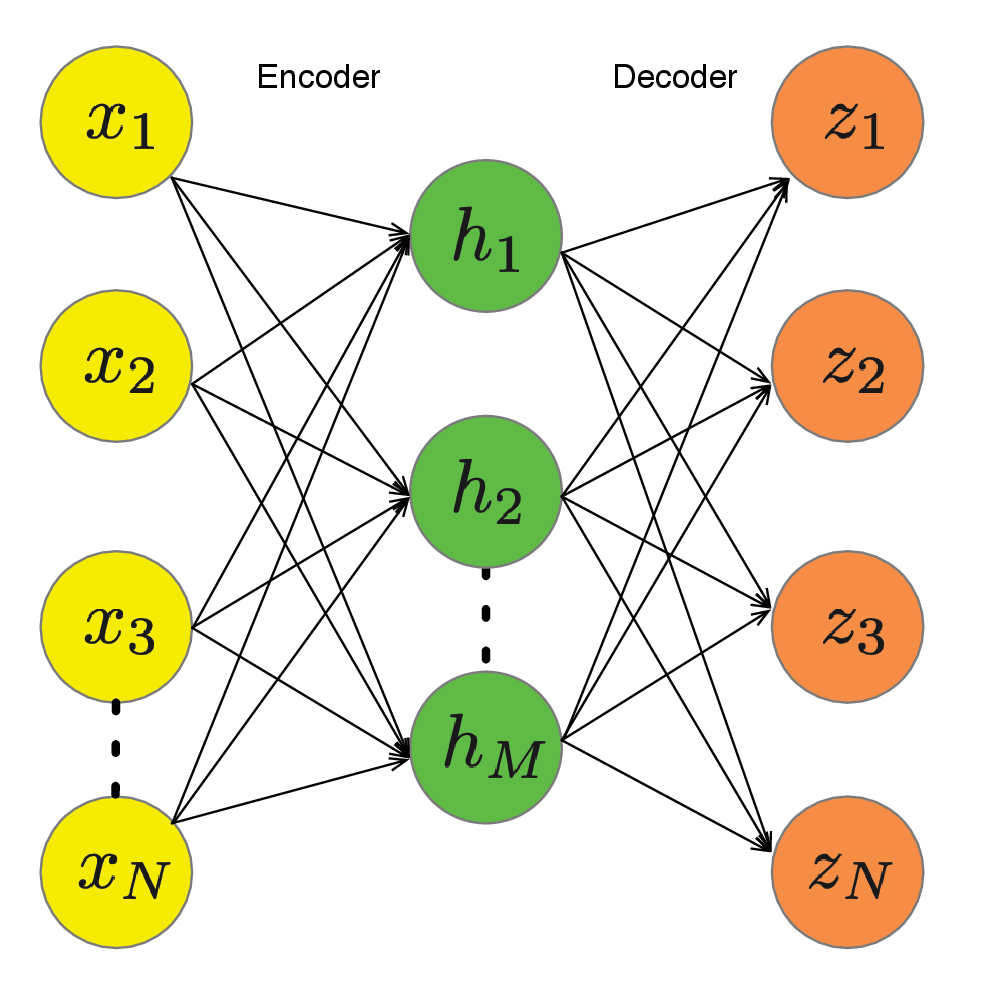}
\caption{Structure of AE with three layers (adapted from \cite{mao2019new}).}
\label{fig:ae}
\end{figure}
An Auto-Encoder (AE) is a neural network model that uses a function to map input data into their short/compressed version subsequently decoded into a closest version of the original input. As shown in Fig. \ref{fig:ae}, an AE has three types of layers, input layer, one or more hidden layers and output layer. The structure of an AE can be considered as an encoder which integrated with a decoder. The encoder transforms an input $\mathbf{x}$ to a hidden representation $\mathbf{h}$ by a non-linear activation function $\varphi$:
\begin{equation}
  \mathbf{h}=\varphi(\mathbf{W}\mathbf{x}+\mathbf{b}),
  \label{eq_encode}
\end{equation} 
Then, the decoder maps the hidden representation $\mathbf{h}$ back to the original representation in a similar way:
\begin{equation}
  \mathbf{z}=\varphi(\mathbf{W'}\mathbf{h}+\mathbf{b'}),
  \label{eq_decoder}
\end{equation}
where $\mathbf{W},\mathbf{b},\mathbf{W}',\mathbf{b}'$ are model parameters and will be optimized to minimize the reconstruction error between $\mathbf{z}$ and $\mathbf{x}$. The average reconstruction error over $N$ data samples can be measured by Mean Squared Error (MSE) and the optimization problem can be expressed as:
\begin{equation}
  \min_{\theta}\frac{1}{N}\sum_{i}^{N}(\mathbf{x}_i-\mathbf{z}_i)^2,
  \label{eq_ae}
\end{equation}
where $\mathbf{x}_i$ is the $i$-th sample. If the input data is highly nonlinear, more hidden layers are required to construct the deep AE. Based on the basic AE model, many variants of AE with quite different properties and implementations have been proposed, such as sparse AEs, denoising AEs and stacked AEs.

AE and its deep models are promising methods to learn high-level representation from raw data \cite{jia2018neural,lu2015novel,WANG2018213,zhao2018fault, yuan2017deep}. For example, in order to avoid learning similar features and misclassification, Jia \emph{et al.} \cite{jia2018neural} propose a Local Connection Network (LCN) constructed by normalized sparse AE (NSAE) for automatic feature extraction from raw vibration signals. Especially, a soft orthonormality constraint is added in the cost function to learn dissimilar features. One of the experiments by using raw data from planetary gearbox has illustrated that the testing accuracy can reach 99.43\%, which is much better than PCA+SVM (41.04\%), Stacked SAE+softmax (34.75\%) and SAE+LCN (94.41\%). Lu \emph{et al.} \cite{lu2015novel} employ a basic AE as the feature extractor to obtain a meaningful representation from highly dimensional bearing signal samples. However, raw data with high dimensionality may lead to heavy computation cost and overfitting with huge model parameters. Therefore, multi-domain features can be extracted first from raw data and then fed to AE-based models. Wang \emph{et al.} \cite{WANG2018213} and Zhao \emph{et al.} \cite{zhao2018fault} feed the frequency spectrum of the vibration signals of planetary gearbox to their proposed stacked denoising AEs. In \cite{yuan2017deep}, multi-features are extracted by time domain, frequency domain and time-frequency domain analysis and then fed to two kinds of AEs. In \cite{xu2020extracting}, fast Fourier transformation is used to convert raw time-domain data into frequency-domain data, then a moving window-based stacked auto-encoder with an exponential function, which incorporates a slope local minimum point, is proposed to extract the degradation trends and improve its monotonicity. 

Further, multi-sensory data can be fused via AE-based models. In practice, more than one sensor would be mounted at different positions to acquire a variety of possible fault signals. The statistical characteristics of one signal may vary from another at a different location and time, which not only leads to the difficulty of selecting artificial features, but also increases uncertainty in fault diagnosis and prognosis. In \cite{chen2017multisensor}, Chen \emph{et al.} feed the time-domain and frequency-domain features extracted from different raw bearing vibration signals into multiple two-layer sparse AEs for feature fusion. Experiments carried out on a rotary machine experimental platform show that feature fusion improves data clustering performance greatly. Ma \emph{et al.} \cite{ma2018deep} employ a deep coupling AE (DCAE) to find a joint feature between vibration signals and acoustic signals. Experimental results show that the classification accuracy of the proposed DCAE with fusion on health assessment for gears is 94.3\%, while that of deep AE without fusion can just reach 88.7\% -- 91.3\%.

AE based models can be naturally integrated with kinds of classifiers to tackle the fault diagnosis problems. A most commonly used classifier is ``softmax'' \cite{shao2017novel, 7983338, lu2017fault, li2018unsupervised, yu2019representation}. For example, Shao \emph{et al.} \cite{shao2017novel} design a new AE with a maximum correntropy based loss function, which can eliminate the impact of background noise in the raw rotating machinery signals. The learned features are finally fed into the ``softmax'' classifier for fault classification. The results show that the average testing accuracy of the proposed method is 94.05\%, and it is much higher than standard deep AE (83.75\%), back propagation (BP) neural network (46.00\%) and SVM (55.75\%). Besides ``softmax'', SVM \cite{lv2017weighted, sun2016sparse}, Support Vector Data Description (SVDD) \cite{mao2019new}, Extreme Learning Machine (ELM) \cite{haidong2018intelligent, roy2018stacked}, RF \cite{thirukovalluru2016generating}, and DNN \cite{sun2016sparse} are usually employed along with AE-based models for fault diagnosis. In \cite{sun2016sparse}, features are learned by sparse AE (SAE) and then used to train a neural network classifier for identifying induction motor faults. In the experiments, two additional classifiers, i.e., SVM and logistic regression (LR), built on top of the SAE, are applied as baseline approaches. The results show that the classification accuracy of SAE with DNN classifier can reach 97.61\%, which is slightly higher than SAE+SVM (96.42\%) and SAE+LR (92.75\%). To accelerate the training speed, Shao \emph{et al.} \cite{haidong2018intelligent} apply ELM as a classifier for intelligent fault diagnosis of rolling bearing. The testing accuracy of ELM is 95.20\%, while that of ``softmax'' is 95.03\% and SVM is 92.80\%. Although these three classifiers achieve basically satisfactory and similar diagnosis results (over 90\%), the training speed of ELM is around 30 and 17 times faster than SVM and ``softmax'', respectively. 

\begin{figure}[htbp]
\centering
\includegraphics[width=3.3in]{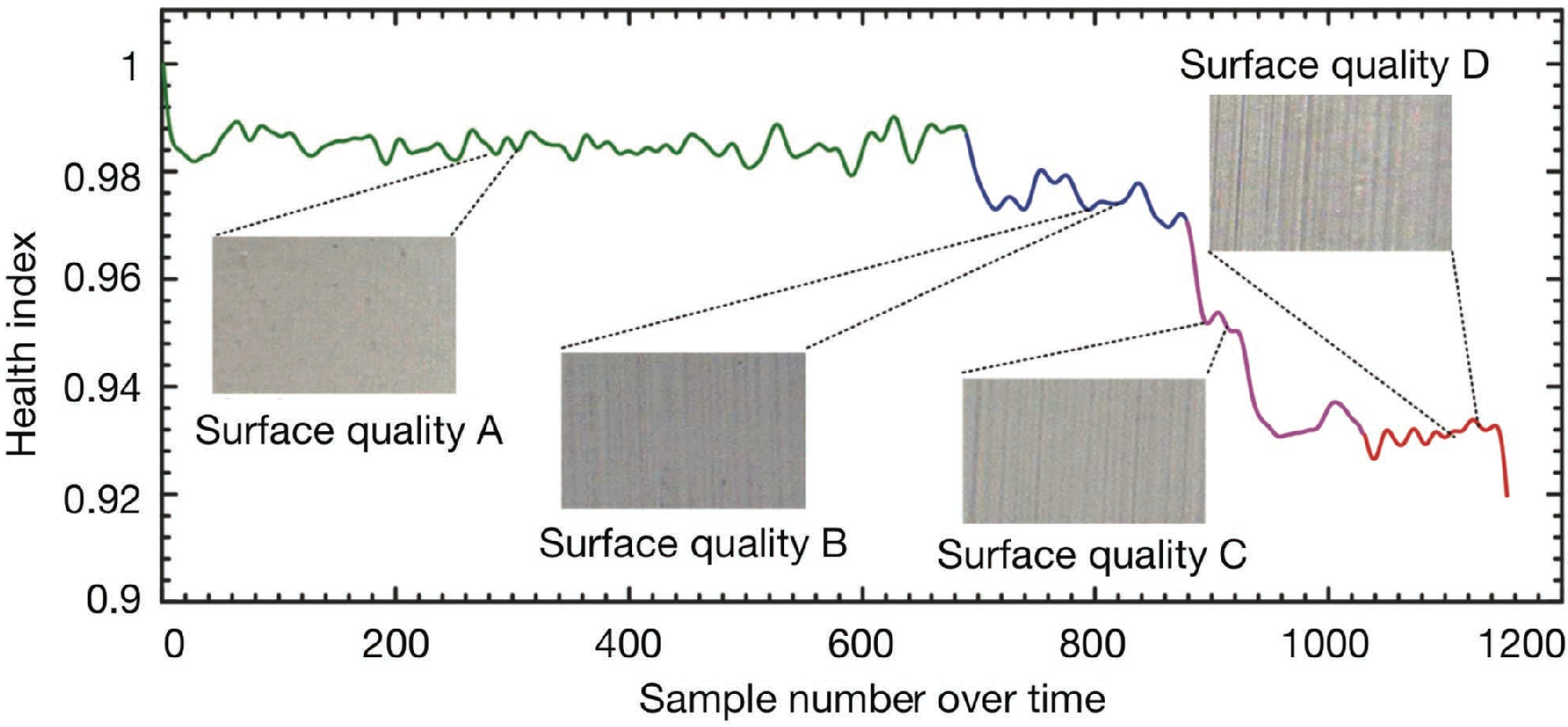}
\caption{Health index based on the similarities of the feature vectors over time \cite{luo2019early}.}
\label{fig:h_index}
\end{figure}

 
Due to the lack of failure history data or labeled data, AE-based models have a great attraction to the degradation process estimation. Typically, these approaches are able to measure the distance to the healthy system condition and also to distinguish different degrees of fault severity. Luo \emph{et al.} \cite{luo2019early} propose a deep AE model to recognize and select the impulse responses from the long-term vibration data, then employ a health index based on Cosine distance to calculate the similarity between the dynamic feature vectors and indicate the slow and gradual degradation process of the machine tool. As shown in Fig. \ref{fig:h_index}, there are four stages in the health index curve: health, slight deterioration, rapid deterioration, and severe deterioration. The early faults can be detected when the cosine similarity value between the standard vector and current vector decreases a lot. A similar approach is proposed in \cite{michau2018feature, michau2018data} to assess the health of a system. However, Michau \emph{et al.} use hierarchical extreme learning machines (HELM), which consists of stacking sparse AEs, to aggregate the features in a single indicator, representing the health of the system. In \cite{wen2018degradation}, Wen \emph{et al.} construct a health indicator from raw data with the variational AE. Then, a degradation estimation model based on kernel density estimation is utilized to identify the deterioration at the early stage of the ball screw. This approach also requires no empirical information and failure history data. Similarly, Lin \emph{et al.} \cite{lin2019novel} employ ensemble stacked AEs to extract features from the fast Fourier transform (FFT) results of raw vibration signals, and use a deep neural network to map the extracted features to a 1-D health indicator ranging from 0 to 1.  

Further, AE-based models are usually combined with various regression models to predict RUL of machinery equipment \cite{xia2018two, ren2018remaining, ma2018predicting, yan2018industrial}. Xia \emph{et al.} \cite{xia2018two} develop a two-stage DNN-based prognosis approach. First, a denoising AE with ``softmax'' is applied to classify the acquired signals of the monitored bearings into different degradation stages. Then, regression models based on ANN are constructed for each health stage. The final RUL result is estimated by smoothing the regression results from different models via the following equation:
\begin{equation}
  RUL(X) = \sum_{i=1}^nP(S_i|X)\cdot R_i(X),
  \label{eq:cost}
\end{equation}
where $X$ denotes the monitored signals, $n$ is the total number of health stages, $P(S_i|X)$ indicates the probability that the sample is classified into the $i$-th class and $R_i(X)$ is the intermediate RUL estimated by the $i$-th ANN regression model. Ren \emph{et al.} \cite{ren2018remaining} propose a similar approach to predict RUL of Lithium-Ion battery. First, a multi-dimensional feature extraction method with AE model is applied to represent battery health degradation, then the RUL prediction model-based DNN is trained for multi-battery remaining cycle life estimation. The experimental results show that the prediction accuracy of the proposed approach is 93.34\%, which is higher than the Bayesian regression model (89.08\%), the SVM model (89.34\%) and the linear regression model (88\%). In \cite{ma2018predicting}, the authors employ a stacked sparse AE and logistic regression to predict the RUL of an aircraft engine. The stacked sparse AE is applied to automatically extract and fuse degradation features from multiple sensors installed on the aircraft engine, while logistic regression is responsible for predicting the RUL. In \cite{yan2018industrial}, Yan \emph{et al.} present a concept of device electrocardiogram (DECG) and propose a RUL prediction methodology called integrated deep denoising AE (IDDA). The IDDA consists of two DDA and a linear regression analysis. The experimental results show that the error between predicted and true values is about 20\%. In \cite{wang2020combining}, Wang \emph{et. al} firstly employ an AE trained with normal data to extract degradation curves of aircraft engines and build a degradation model template library. Then, they compare each test object with all template curves to get similarities and corresponding RULs based on a sliding window and complexity-invariant distance. At last, the estimated RUL can be obtained by calculating the weighted average of highly relevant corresponding RULs.

\subsection{Convolutional Neural Network (CNN)}
\begin{figure*}[htbp]
\centering
\includegraphics[width=6.5in]{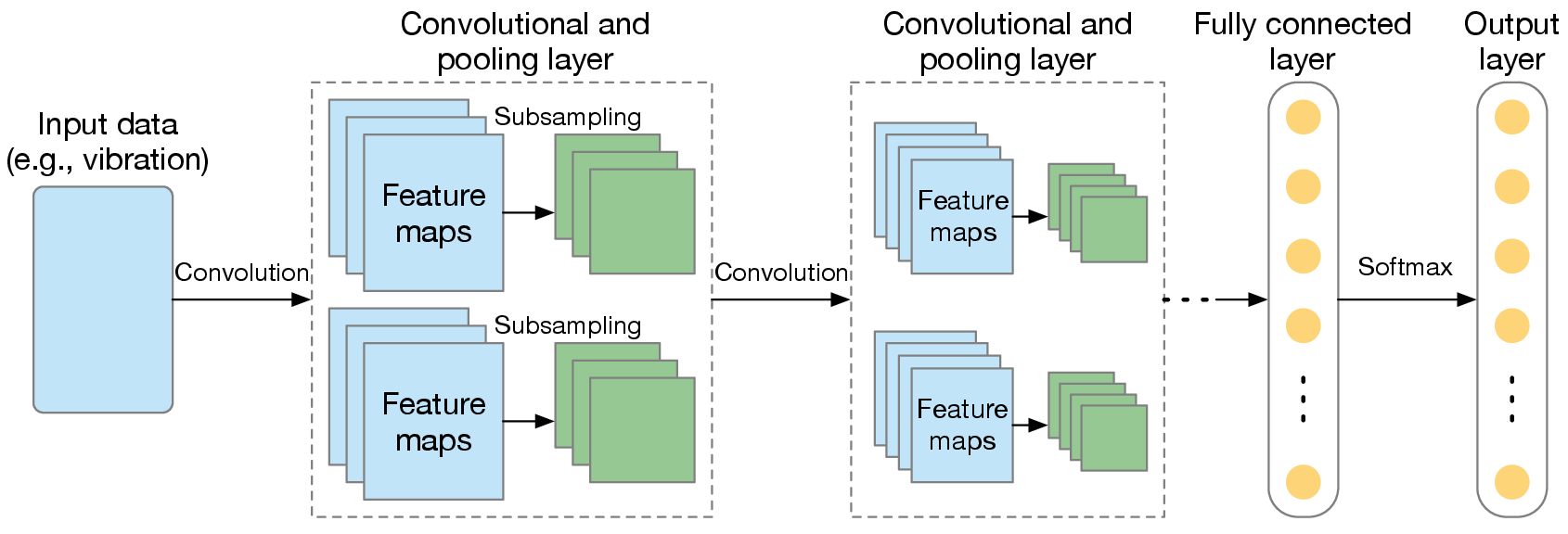}
\caption{A typical architecture of convolutional neural networks (CNN). (adapted from \cite{jing2017convolutional}).}
\label{fig:cnn_arch}
\end{figure*}

Convolutional Neural Network (CNN) is one of the most notable deep learning models due to its shared weights and ability of local field representation \cite{lecun2015deep, lecun1990handwritten}. CNN can extract the local features of the input data and combine them layer by layer to generate high-level features. As illustrated in Fig. \ref{fig:cnn_arch}, a typical CNN structure basically consists of input layer, convolution layer, pooling layer, and fully connected layer. 

\noindent{\textbf{Input layer}}: The input layer can be presented in a two-dimensional manner such as time-frequency spectrum or a one-dimensional manner such as time-series data, e.g., the input data can be represented as $X \in R^{A\times B}$, where $A$ and $B$ are the dimensions of the input data.

\noindent{\textbf{Convolution layer}}: In the convolution layer, the convolution kernel (filter) convolutes the input data from the previous layer through a set of weights and composes a feature output, generally called as a feature map. The output of the convolutional layer can be calculated as:
\begin{equation}
  Y_{cn}=f(X*W_{cn}+b_{cn}),
  \label{eq_cnn}
\end{equation}
where ``*'' represents an operator of convolution, $cn$ denotes the number of convolution filters, $W_{cn}$ is the weight matrix of $cn$-th filter kernel, $b_{cn}$ is the filter kernel bias and $f$ is an activation function such as rectified linear units (ReLU).

\noindent{\textbf{Pooling layer}}: The essence of pooling operation is sampling, which is used to reduce model parameters and retain effective information. At the same time, overfitting can be avoided in some extent and training speed can be improved. The most commonly used pooling layer is the max-pooling layer, which can extract max value from $Y_{cn}$ as follows:
\begin{equation}
  P_{cn}= \max_{S^{M\times N}}(Y_{cn}),
  \label{eq_pooling}
\end{equation}
where $S^{M\times N}$ is a scale matrix of pooling, $M$ and $N$ are the dimension of $S$.

\noindent{\textbf{Fully connected layer}}: After several combination forms of convolution layer and pooling layer, multiple fully-connected layers will follow, which can convert the matrix in filter to a column or a row. Finally, a classification or regression layer can be added to achieve specific aims.

In the field of PdM, CNN has shown dramatic capability in extracting useful and robust features from monitoring data. For one-dimension (1D) monitoring signals, Qin \emph{et al.} \cite{qin2019rolling} build an end to end 1D-CNN that reflects the raw vibration signals to fault types. The result shows that the proposed model can achieve about 99\% accuracy through hyper parameters tuning. In \cite{liu2018real}, Liu \emph{et al.} develop a novel shallow 1-D CNN fault diagnosis model (CNNDM-1D) using raw vibration signal. TCNNDM-1D comprises only two convolutional layers, two pooling layers and one full-connected layer, and fuses feature learning and health classification into a single body. The experimental result shows that the accuracy of CNNDM-1D can reach 99.886\%. Kiranyaz \emph{et al.} \cite{kiranyaz2018real} propose a real-time and highly accurate modular multilevel converter (MMC) circuit monitoring system for early fault detection and identification leveraging adaptive 1D CNN. The proposed approach is directly applicable to the raw voltage and thus eliminates the need for any feature extraction algorithm. Simulation results obtained using a 4-cell, 8-switch MMC topology demonstrate that the proposed system has high reliability to avoid any false alarm and achieves a detection probability of 0.989, and average identification probability of 0.997 in less than 100 ms.

CNN combined with a certain signal processing algorithm is often adopted for a better fault diagnosis performance. Although 1D CNN requires a limited amount of data for effectively training and has low-cost hardware implementation for real-time applications, CNN has poor feature extraction capability for sensor data with 1D format. To address this issue, Li \emph{et al.} \cite{li2019sensor} propose a novel sensor data-driven fault diagnosis method, named ST-CNN, by combining S-transform (ST) algorithm and CNN. The ST layer automatically converts the sensor data into 2D time-frequency matrix without manual conversion work. Chen \emph{et al.} \cite{chen2015gearbox} extract a total of 256 statistic features of each gear failure sample from time and frequency domains, and then reshaped them into a matrix (16 $\times$ 16) as the input of the convolution layer, which shows better classification performance in comparison with SVM. Wang \emph{et al.} \cite{wang2019deep} convert a raw acceleration signal to a uniform-sized time-frequency image which then is fed into a universal bearing fault diagnosis model transferred from a well-known Alexnet model. In stead of converting 1D data into 2D format, CNN can learn features from 2-D input more effectively \cite{oh2019convolutional}. Jia \emph{et al.} \cite{jia2019rotating} use BoW model to extract features from infrared thermography images of rotating machinery to implement the automatic fault diagnosis. Liu \emph{et al.} \cite{liu2017infrared} use infrared images for rotating machinery monitoring and fault diagnosis.  Peng \emph{et. al} \cite{peng2020multibranch} proposes a novel multi-branch and multi-scale convolutional neural network that can automatically learn and fuse abundant and complementary fault information from the multiple signal components and time scales of the vibration signals.

Health Indicator (HI) or degradation process of machinery equipment can be constructed via CNN-based models for fault prognosis. The existing manual HI construction methods usually need prior knowledge of experts to design feature extraction and data fusion algorithms. In order to handle this issue, Guo \emph{et al.} \cite{guo2018machinery} propose a CNN-based method to automatically learn features and construct an HI. First, several convolution and pooling operations are stacked to learn features, and then these learned features are mapped into an HI through a nonlinear mapping operation. Second, the performance of the HI is further improved by detecting and removing outlier regions. The experimental results show that the CNN-based HI provides the best results in terms of trendability, monotonicity and scale similarity compared with other HIs based on stacked AE, self-organizing map and fully-connected neural network. Yoo \emph{et al.} \cite{yoo2018novel} propose a novel time-frequency image feature to construct HI. To convert the 1D vibration signals to a 2D image, the continuous wavelet transform (CWT) extracts the time-frequency image features, i.e., the wavelet power spectrum. Then, the obtained image features are fed into a 2D CNN to construct the HI. Cheng \emph{et al.} \cite{cheng2018online} train a novel CNN model to successfully extract a novel nonlinear degradation energy index (DEI) to describe the degradation trend of the training bearing, according to the natural frequencies of bearing components. Then, a $\epsilon$-SVR model is introduced so that the evolution of the degradation can be forecast till the bearing failure.

\begin{figure}[htbp]
\centering
\includegraphics[width=3.5in]{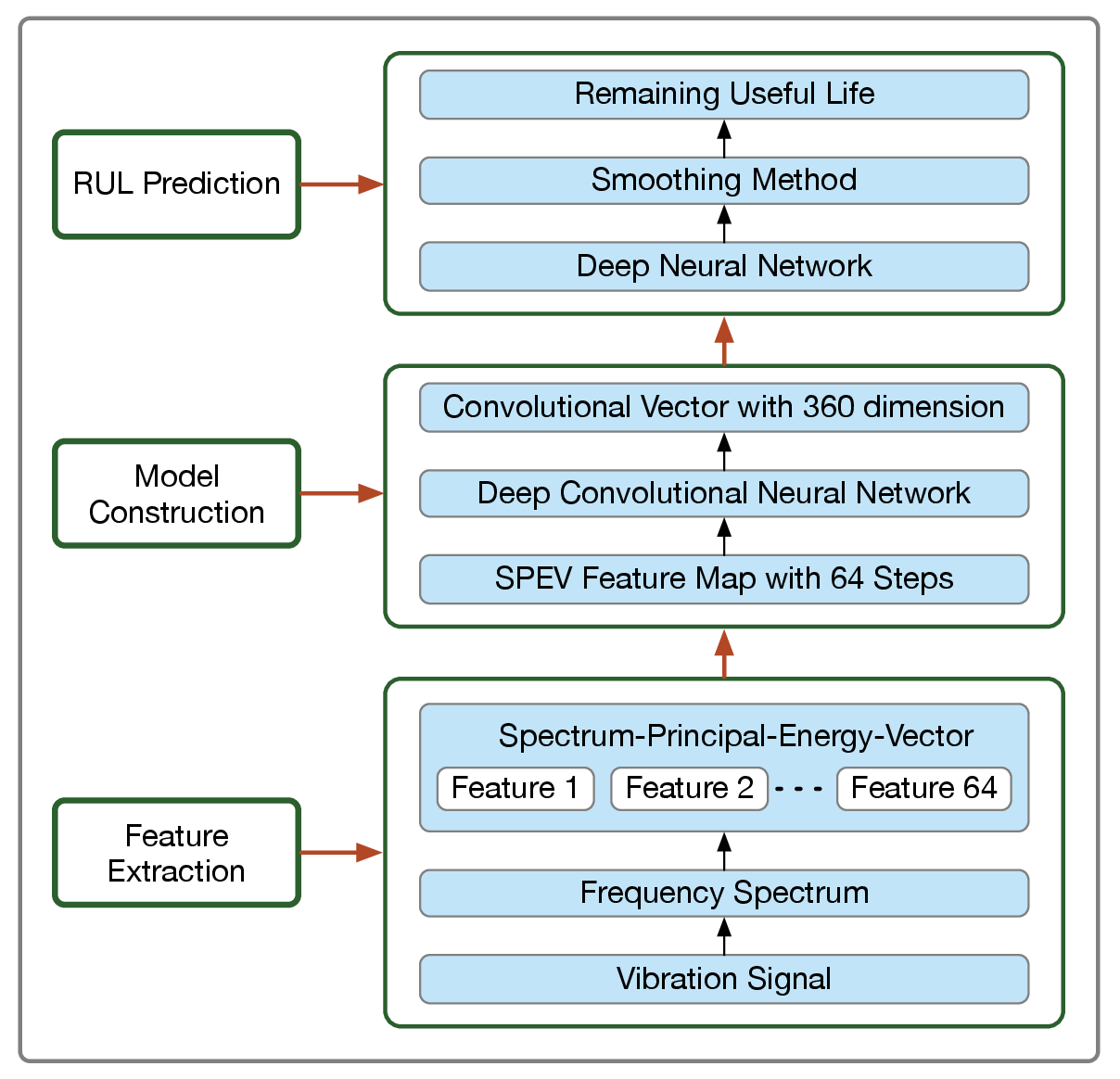}
\caption{A framework of RUL prediction by applying CNN \cite{ren2018prediction}.}
\label{fig:rul_arch}
\end{figure}

Furthermore, the applications of CNN on RUL prediction have been widely investigated. Ren \emph{et al.} \cite{ren2018prediction} construct a CNN combined with a smoothing method for bearing RUL prediction. As shown in Fig. \ref{fig:rul_arch}, the vibration signal is converted to discrete frequency spectrum (named the spectrum-principal-energy-vector) via fast Fourier transform (FFT), then the deep CNN analyzes this spectrum-principal-energy-vector and obtains a series of eigenvectors. Afterwards, the deep neural network model is used for regression prediction to obtain the RUL of the bearing. In addition, this paper uses the forward prediction data to linearly smooth the current forecast data to alleviate the problem of discontinuous predicted RUL. The results showed that the proposed method can significantly improve the prediction accuracy of bearing RUL. Babu \emph{et al.} \cite{babu2016deep} build a 2D deep CNN to predict the RUL of system based on normalized variate time series from sensor signals, where one dimension of the 2D input is the number of sensors. Average pooling is adopted in their work and a linear regression layer is placed on the top layer. In this study, the CNN structure is employed to extract the local data features through the deep learning network for better prognostics. In \cite{li2019deep}, the time-frequency domain information is obtained by applying short-time Fourier transform and explored for prognostics, and multi-scale feature extraction is implemented using CNN. Experiments on a popular rolling bearing dataset collected from the PRONOSTIA platform are carried out to show the effectiveness and high accuracy of the proposed method. In \cite{zhu2018estimation}, Zhu \emph{et al.} first derive Time Frequency Representation (TFR) of each sample by using wavelet transform, then feed the TFR to a Multi-Scale CNN (MSCNN) to extract more identifiable features. Finally, a regression layer following MSCNN to perform RUL estimation. Other CNN variants are also applied to predict RUL (e.g.: deep CNN \cite{8982045}, double CNN \cite{yang2019remaining}, residual CNN \cite{wen2019new}).

\begin{figure}[htbp]
\centering
\includegraphics[width=3.5in]{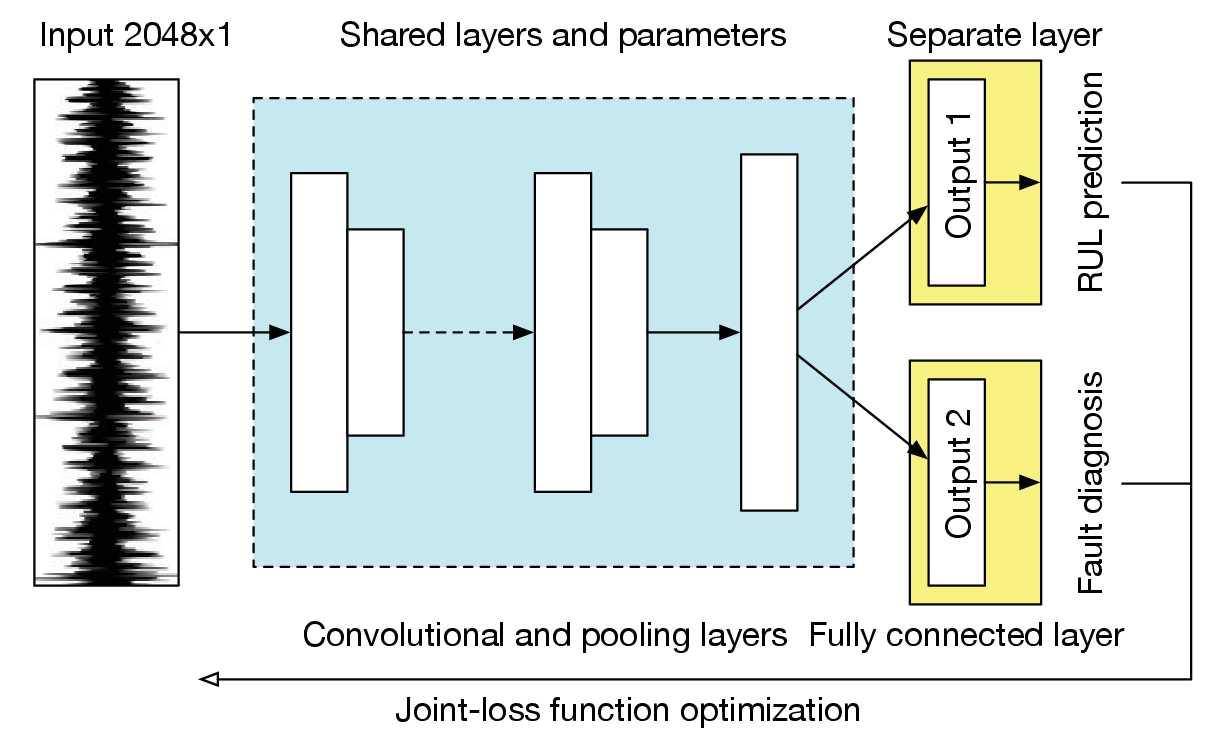}
\caption{The JL-CNN architecture (adapted from \cite{liu2019simultaneous}).}
\label{fig:joint_cnn}
\end{figure}

Previously, fault diagnosis and RUL prediction are always been investigated separately, however, some information about these two tasks can be shared to improve the performance. Therefore, Liu \emph{et al.} \cite{liu2019simultaneous} propose a joint-loss CNN (JL-CNN) architecture to capture the common features between fault diagnosis and RUL prediction. As shown in Fig. \ref{fig:joint_cnn}, JL-CNN has a shared CNN and two independent fully connected networks. Due to the shared network, the feature representations of one task can be also applied by another task, which can lead to co-learning. On the other hand, the independent fully connected networks can achieve specific targets (i.e.: fault diagnosis and RUL prediction). To train such a hybrid network, a joint loss function is constructed as:
\begin{equation}
  J(W) = J_1(W)+\lambda J_2(W),
  \label{eq_joint_loss}
\end{equation}
where $J_1(W)$ and $J_2(W)$ denote the loss function of RUL prediction and fault diagnosis respectively, $\lambda$ is a penalty factor for controlling the weight of the two tasks. The experimental results show that the MSE of the proposed method decreases 82.7\% and 24.9\% respectively compared with SVR and traditional CNN. 

\subsection{Recurrent Neural Network (RNN)}
\begin{figure}[htbp]
\centering
\subfigure[The architecture of RNN across a time step.]{
    	\label{fig_rnn_a} 
    	\includegraphics[width=2in]{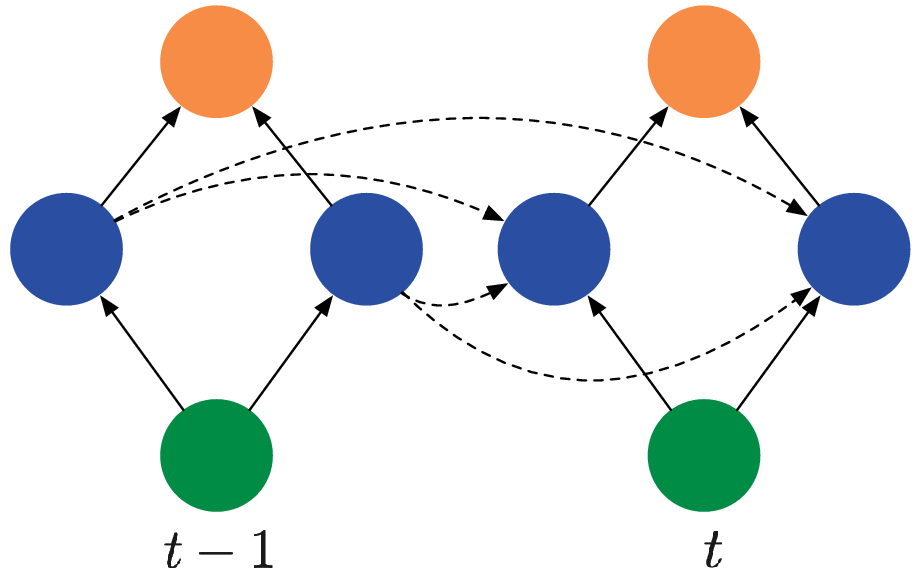}}
\subfigure[The architecture of a LSTM memory cell.]{
	\label{fig_rnn_b} 
     	\includegraphics[width=3in]{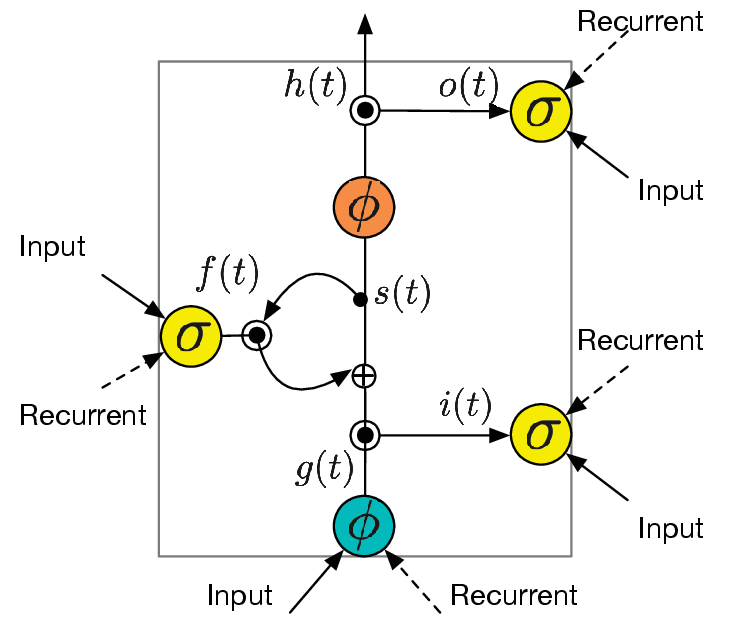}}
\caption{The basic architecture of RNN and LSTM \cite{guo2017recurrent}. (a) The architecture of RNN across a time step. (b) The architecture of a LSTM memory cell.}
\label{fig_rnn} 
\vspace{-0.15in}
\end{figure}

Recurrent Neural Networks (RNNs) are a group of neural networks for dealing with sequential data. As a sequential model, RNN can build cycle connections among its hidden units and keep a memory of previous inputs in the network's internal state. As shown in Fig. \ref{fig_rnn_a}, the transition function $\mathbb{H}$ defined in each time step $t$ takes the current time information $x_t$ and the previous hidden output $h_{t−1}$ and updates the current hidden output as follows:
\begin{equation}
  h_t = \mathbb{H}(x_t,h_{t-1}).
  \label{eq_rnn}
\end{equation}
The final hidden output at the last time step is the learned representation of the whole input sequential data. However, due to being trained with Back Propagation Through Time (BPTT), RNN always has the notorious gradient vanishing/exploding issue. In order to overcome this issue, Long Short-Term Memory (LSTM) and Gated Recurrent Units (GRU) are proposed. Specifically, as shown in Fig. \ref{fig_rnn_b}, LSTM is enhanced by adding ``forget'' gates and has shown marvelous capability in memorizing and modeling the long-term dependency in data. LSTM is one of the most commonly used models when working with time-dependent data. 

RNN has been widely used in fault diagnosis in recent years since it significantly outperforms other network structures in sequence learning problems. In \cite{li2018intelligent}, Li \emph{et al.} propose a deep RNN (DRNN) for rotating machinery fault diagnosis. The DRNN is constructed by the stacks of the recurrent hidden layer to automatically extract the features from frequency signal sequences, and the ``softmax'' classifier is employed for rotating machinery fault recognition. In \cite{yuan2019intelligent}, Yuan \emph{et al.} propose a three-stage fault diagnosis approach based on GRU. This proposed approach splits the raw data into several sequence units as the input of GRU. GRU is built to extract the dynamic features from the sequence units effectively and trained through batch normalization algorithm to reduce the influence of the covariate displacement, Finally, ``softmax'' is used to classify the faults. Similarly, Zhao \emph{et al.} \cite{zhao2019intelligent} also use GRU for fault diagnosis of rolling bearing. However, artificial fish swarm algorithm is applied to obtain the key parameters of deep GRU, and an extreme learning machine is employed to replace the ``softmax'' classifier to achieve better diagnostic results. Besides, LSTM is also employed for fault diagnosis. In \cite{yang2018rotating}, both spatial and temporal dependencies are handled by LSTM to identify rotating machinery faults based on the measurement signals from multiple sensors. In \cite{zhao2018sequential}, Zhao \emph{et al.} provide an end-to-end framework based on batch-normalization-based LSTM to learn the representation of raw input data and classifier simultaneously without taking the conventional ``feature + classifier'' strategy. The proposed method is evaluated in the Tennessee Eastman benchmark process and the results show that LSTM can better separate different faults and provide more promising fault diagnosis performance. 

Due to the remarkable ability on long-term memory and time-dependent data, many efforts have been devoted to RNN-based (especially LSTM-based) RUL prediction. In \cite{chen2019gated}, Chen \emph{et al.} employ kernel principal component analysis (KPCA) to extract nonlinear feature and then used a GRU-based RNN to predict RUL. The effectiveness of the proposed approach for RUL prediction of a nonlinear degradation process is evaluated by a case study of commercial modular aero-propulsion system simulation data (C-MAPSS-Data) from NASA. In \cite{chen2020machine}, Chen \emph{et. al}  Honga \emph{et al.} \cite{hong2019fault} combine LSTM method for the first time with the voltage abnormality prediction of the battery system. Given the amount of data acquired from the service and management center for electric vehicles (SMC-EV) in Beijing, LSTM network is applied to perform battery voltage prediction for all-climate electric vehicles. Wu \emph{et al.} \cite{wu2018remaining} employ SVM to detect the starting time of degradation and used vanilla LSTM to predict RUL. However, the RUL requires labeling at every time step for each sample. In addition, an appropriate threshold needs to be defined in advance when employing SVM. In \cite{wu2018approach}, the features are first selected out using correlation metric and monotonicity metric, and then fed into LSTM networks with a concatenated one-hot vector. Finally, a fully-connected layer is applied to map the hidden state into the parameters for sampling consequent point. Different from \cite{wu2018remaining}, all prediction tasks can be completed without any pre-defined threshold or machine learning methods. Miao \emph{et al.} \cite{miao2019joint} propose dual-task deep LSTM networks to perform the RUL prediction and degradation stage (DS) assessment of aero-engines in a parallel way. The experimental results based on the dataset C-MAPSS show a relatively satisfactory result in terms of both DS assessment and the RUL prediction. 

An important task in RUL estimation is the construction of a suitable health indicator (HI) to infer the health condition. Guo \emph{et al.} \cite{guo2017recurrent} propose an RNN based Health Indicator (RNN-HI) to predict the RUL of bearings with LSTM cells used in RNN layers. With monotonicity and correlation metrics, the most sensitive features are selected from an original feature set and then fed into an RNN network to construct the RNN-HI, from which RUL is estimated. With experiments on generator bearings from wind turbines, the proposed RNN-HI is illustrated to achieve better performance than a self-organizing map (SOM) based method. However, the RUL is calculated through an exponential model with pre-set failure threshold of RNN-HI rather than through the trained RNN directly, which means that the advantage of the LSTM cell is not fully utilized. In \cite{ning2018feature}, Ning \emph{et al.} also implement RNN to fuse the selected sensitive features to construct an HI. This HI incorporates mutual information of multiple features and is correlated with the damage and degradation of bearing. Chen \emph{et.al } \cite{chen2020health} propose a convolutional RNN model for health indicator construction, which extracts locally sequential features directly in convolution feature maps through CNN, and further embeds an in-network RNN layer that connects these sequential features seamlessly. The finally obtained features in RNN layer have global receptive field and encode time-series information.
 
\subsection{Deep Belief Networks (DBN) }
\begin{figure}[htbp]
\centering
\includegraphics[width=2.5in]{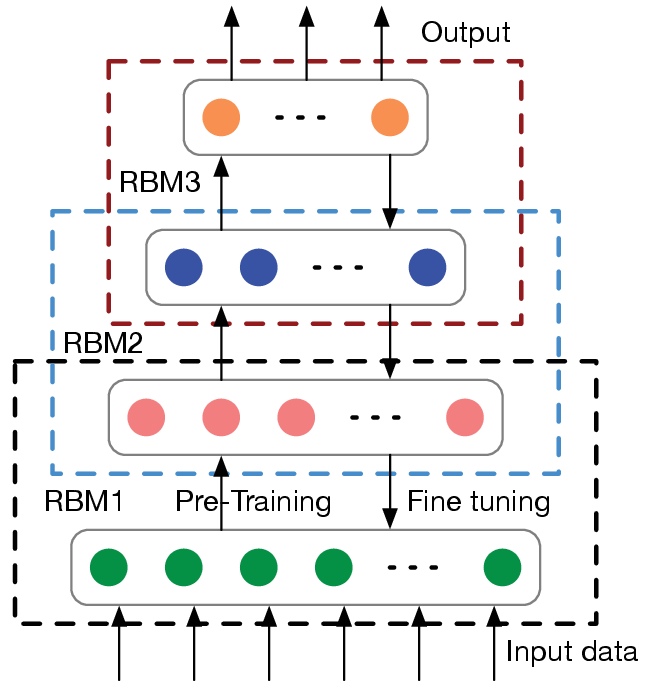}
\caption{The structure of a 3-layer DNB \cite{niu2018fault}.}
\label{fig:dbn_arch}
\end{figure}

Deep belief network (DBN) can be constructed by stacking multiple restricted Boltzmann machines (RBMs), where the output of the $l$-th layer (hidden units) is used as the input of the $(l+1)$-th layer (visible units). DBN can be trained in a greedy layer-wise unsupervised way. After pre-training, the parameters of this deep architecture can be further fine-tuned with respect to a proxy for the DBN log-likelihood, or with respect to labels of training data by adding a ``softmax'' layer as the top layer, which is shown in Fig. \ref{fig:dbn_arch}.  

Similar to CNN and AE, DBN also can be employed to extract high-level features from monitoring signals. In the work of \cite{liang2018bearing}, data from two accelerometers mounted on the load end and fan end are processed by multiple DBNs for feature extraction. then the faulty conditions based on each extracted feature are determined with ``softmax'', and the final health condition is fused by DS evidence theory. An accuracy of 98.8\% is accomplished while including the load change from 1 hp to 2 and 3 hp. In contrast, the accuracy of SAE suffers the most from this load change, and the accuracy employing CNN is also lower than DBN. Pan \emph{et al.} \cite{pan2018intelligent} develop an intelligent fault diagnosis method using DBN for rolling bearing fault identification. In this method, discrete wavelet packet transform is first used to calculate the original features from raw vibration signals. Due to information redundancy of the original features, a DBN with three hidden layers for deep-layer-wise feature extraction is applied for dimensionality reduction. Zhang \emph{et al.} \cite{zhang2018analog} propose a novel analog circuit incipient fault diagnosis method using DBN-based feature extraction. In the diagnosis scheme, DBN method has been used to extract the deep and intrinsic features from the measured time responses, where the learning rates of DBN have been produced by using an algorithm named QPSO. An SVM based incipient fault diagnosis model is set up to classify different incipient fault classes. Shen \emph{et al.} \cite{shen2019improved} propose an improved hierarchical adaptive DBN for bearing fault type and degree diagnosis, where DBN is enhanced by Nesterov momentum and a learning rate adjustment strategy. The improved DBN is applied to directly extract deep data features from the frequency spectrum in stead of the manually extracted features. The ``softmax'' classifier is connected to the top of DBN as the classification layer.

In addition to being used to extract features, DBN can also be applied as a classifier (without additional classifier, e.g. ``softmax'') for fault classification and identification. Tamilselvan \emph{et al.} \cite{tamilselvan2013failure} propose a multi-sensory DBN-based health state classification model. The model was verified in benchmark classification problems and two health diagnosis applications including aircraft engine health diagnosis and electric power transformer health diagnosis. Shao \emph{et al.} propose an adaptive DBN and dual-tree complex wavelet packet (DTCWPT) \cite{shao2017rolling}. The DTCWPT first prepossesses the vibration signals, where an original feature set with $9×8$ feature parameters is generated. The decomposition level is $3$, and the db5 function, which defines the scaling coefficients of the Daubechies wavelet, is taken as the basis function. Then a $5$-layer adaptive DBN of $(72, 400, 250, 100, 16)$ structure is applied for bearing fault classification. The average accuracy is 94.38\%, which is much better compared to convention ANN (61.13\%), GRNN (69.38\%) and SVM (66.88\%). In \cite{chen2017multisensor}, Chen \emph{et al.} employ sparse AE to fuse the time-domain and frequency-domain features, and then the fused features were utilized to train the DBN based classification model. In \cite{zhu2019novel}, PCA technique is adopted to reduce the dimension of raw bearing vibration signals and extract the bearing fault features in terms of primary eigenvalues and eigenvectors. Then, a DBN is trained for fault classification and diagnosis. The experimental results demonstrate the effectiveness of the PCA-DBN based fault diagnosis approach with a more than 90\% accuracy rate. Wang \emph{et al.} \cite{wang2018data} present an DBN-based method to detect multiple faults in axial piston pumps. Data indicators are extracted from the time, frequency and time-frequency domain raw signals and fed into DBNs to classify the multiple faults in axial piston pumps. The fault classification accuracy are 99.17\%, and 97.40\%, respectively, for benchmark data with 6 classes of bearing faults and for experimental test rigs with 5 classes of axial piston pump faults.

Many efforts also have been devoted to using DBN for RUL estimation and early fault detection. In \cite{deutsch2017using}, a DBN-feedforward neural network (FNN) is applied to perform automatic self-taught feature learning with DBN and RUL prediction with FNN. Two accelerometers were mounted on the bearing housing, in directions perpendicular to the shaft, and data is collected every 5 min, with a 102.4 kHz sampling frequency, and a duration of 2 seconds. Experimental results demonstrate the proposed DBN based approach can accurately predict the true RUL 5 min and 50 min into the future. Li \emph{et al.} \cite{li2018lithium} propose a deep belief networks (DBN) method for lithium-ion battery RUL prediction. The proposed method is trained with historical battery capacity data. With the powerful fitting ability of DBN, the proposed method can track capacity degradation and predict the RUL. The experimental results show that the proposed method has high accuracy in capacity fade prediction and RUL prediction. In \cite{wang2019early}, Wang \emph{et al.} utilize a multi-operation condition partition scheme to segment normal state data of wind turbines into multiple different clusters, and then constructed an optimized DBN (ODBN) model with chicken swarm optimization to capture the normal behavior in each cluster. Finally, Mahalanobis distance measure is employed to identify the early anomalies that occur in the operation of the wind turbines. In \cite{zhao2017lithium}, the authors apply DBN to extract features from the capacity degradation of lithium-ion batteries, and fed the extracted features to a relevance vector machine (RVM) to provide RUL prediction. Extensive experiments are conducted based on the CALCE battery datasets and the results show that, compared with standard DBN and RVM, the proposed method has higher accuracy.

\subsection{Generative Adversarial Network (GAN)}
\begin{figure}[htbp]
\centering
\includegraphics[width=3.5in]{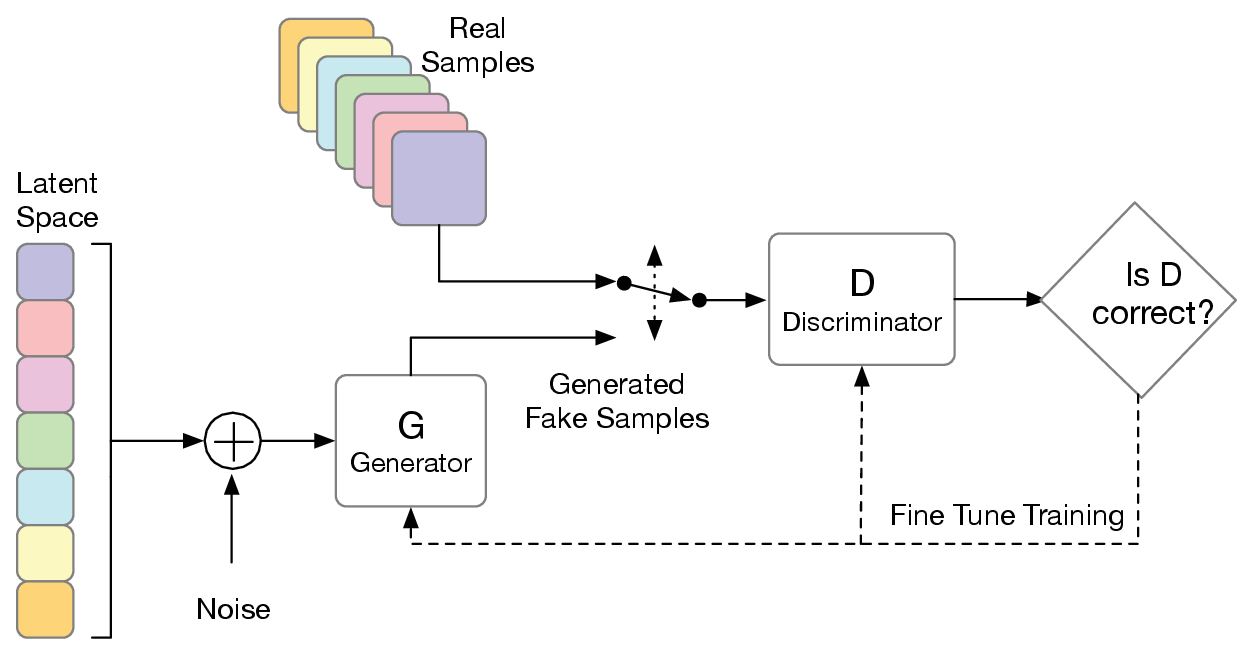}
\caption{The structure of GAN.}
\label{fig:gan_arch}
\end{figure}

Generative adversarial network (GAN) was initially introduced by Goodfellow \emph{et al.} \cite{goodfellow2014generative} in 2014. A typical GAN framework consists of a generator ($G$) and a discriminator ($D$), as illustrated in Fig. \ref{fig:gan_arch}. The generator ($G$) generates fake samples (e.g., time series with sequences) from a random latent space as its inputs, and feeds the generated samples to the discriminator ($D$), which will try to distinguish the generated (i.e. “fake”) samples from the original dataset. The concept of GAN is based on the idea of competition, in which $G$ and $D$ are competing to outsmart each other and improve their own capability of imitation and discrimination respectively.

GAN is firstly utilized as a data augmentation technique to address the class imbalance issue in the field of PdM. In \cite{lee2017application}, the authors illustrated that GAN is able to generate adequate oversampled data when an imbalance ratio is minor. Also, a hybrid oversampling method combining adaptive synthetic sampling (ADASYN) with GAN is designed to resolve the inability of the GAN generator to create meaningful data when the original sample data is scarce. In \cite{suh2019generative}, Suh \emph{et al.} propose DCWGAN-GP based on Wasserstein GAN with a gradient penalty (WGAN-GP) and deep convolutional GAN (DCGAN) to address the data imbalance issue for bearing fault detection and diagnosis. Experiments demonstrated that the proposed method improves accuracy by 7.2 and 4.27\% points on average and gives maximum values with 5.97 and 3.57\% points higher accuracy than the original DCGAN approach. Shao \emph{et al.} \cite{shao2019generative} develop an auxiliary classifier GAN (ACGAN)-based framework to learn from mechanical sensor signals and generate realistic one-dimensional raw data. The proposed approach is designed to produce realistic synthesized signals with labels and the generated signals can be used as augmented data for further applications in machine fault diagnosis. There exist some (but rare) efforts devoting on estimating RUL based on GAN. In \cite{wang2019generalization}, Wang \emph{et al.} implement Wasserstein generative adversarial network (WGAN) to generate simulated signals for balancing the training dataset, and fed the real and artificial signals to stacked AEs for fault classification. In \cite{mao2019imbalanced}, Mao \emph{et al.} derive frequency spectrum of fault samples by using fast Fourier transform to pre-process the original vibration signal. Then, the spectrum data is fed into a GAN to generate synthetic minority samples. Finally, the synthetic samples are utilized to train a stacked denoising AE model for fault diagnosis. In \cite{zhou2020deep}, the generator is designed to generate those fault feature extracted from a few fault samples via AE instead of fault data sample. The training of the generator is guided by fault feature and fault diagnosis error instead of the statistical coincidence of traditional GAN. The discriminator is designed to filter the unqualified generated samples for more accurate fault diagnosis.

\begin{figure}[htbp]
\centering
\includegraphics[width=3.5in]{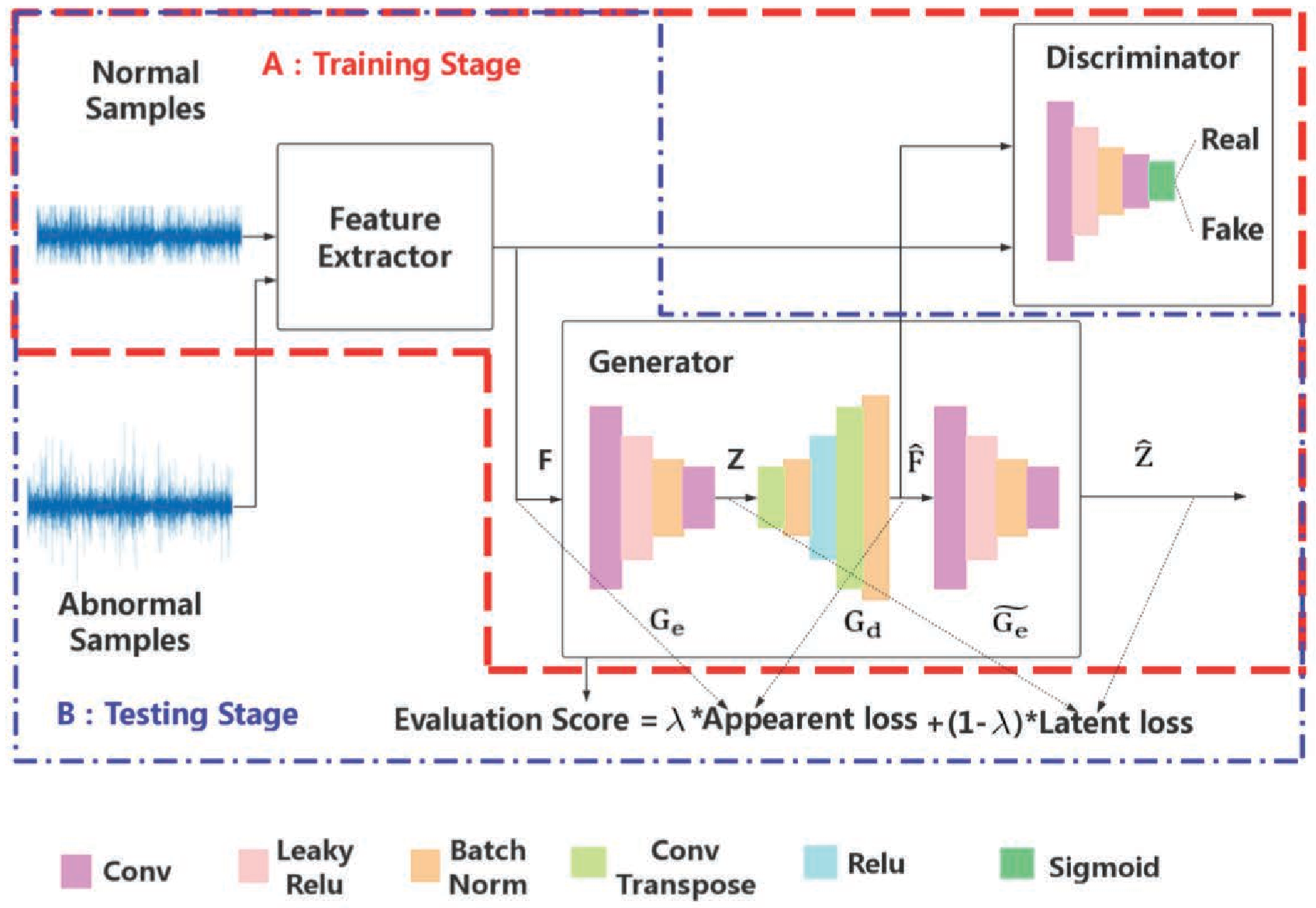}
\caption{Overview of the proposed approach in \cite{jiang2019gan}. The basic network architecture of the generator and discriminator is based on DCGAN. The generator has an encoder-decoder-encoder three-sub-network. In the training stage, only normal samples are involved. In the testing stage, abnormal samples can be discriminated by a higher anomaly score.}
\label{fig:ganomaly}
\end{figure}

Besides generating synthetic samples, GAN also can be directly trained for fault identification. In \cite{akcay2018ganomaly}, Akcay \emph{et al.} propose a generic anomaly detection architecture called GANomaly. GANomaly employs an encoder-decoder-encoder sub-network and three loss functions in the generator to capture distinguishing features in both input images and latent space. At inference time, a larger distance metric from this learned data distribution indicates an anomaly. Jiang \emph{et al.} \cite{jiang2019gan} adopt GANomaly for anomaly detection in the industrial field. A similar encoder-decoder-encoder three-sub-network generator is employed as shown in Fig. \ref{fig:ganomaly}. The generator is trained only using the extracted features from normal samples. Anomaly scores (made up of apparent loss and latent loss) is designed for anomaly detection. Experimental studies based on a benchmark rolling bearing dataset acquired by Case Western Reserve University illustrate that the proposed approach can distinguish abnormal samples from normal samples with 100\% accuracy. Different from GANomaly, in \cite{ding2019generative}, Ding \emph{et al.} present an ensemble GAN approach. This approach uses multiple GANs to learn the data distribution for each health condition and employs a semi-supervised method to enhance the feature extraction ability of the discriminator of each GAN. Finally, a ``softmax'' function is used to ensemble all discriminators for fault diagnosis. The experiments only used 10\% and 20\% of the selected rolling bearing and gearbox datasets as the training data, respectively, and the accuracies were still above 97\% for different working conditions.

For fault prognosis, Khan \emph{et al.} \cite{khan2018towards} employ generative models to model the trend in a bearing's health indicator (HI) and then used those models to generate future trajectories of a bearing's health indicator. These trajectories of a bearing's HI can be used to estimate its RUL by determining the time at which the HI exceeds the failure threshold. Moreover, GANs can be further improved as more and more historical data on bearing degradation becomes available. The proposed method was tested using publicly available run-to-failure test data by IMS, University of Cincinnati. The results clearly indicate the feasibility of using GANs in modeling the degradation behavior of bearings and using them to predict the future values of a bearing's health indicator, which is critical in determining the RUL of a bearing. 

\subsection{Transfer Learning}
\begin{figure}[htbp]
\centering
\includegraphics[width=2.5in]{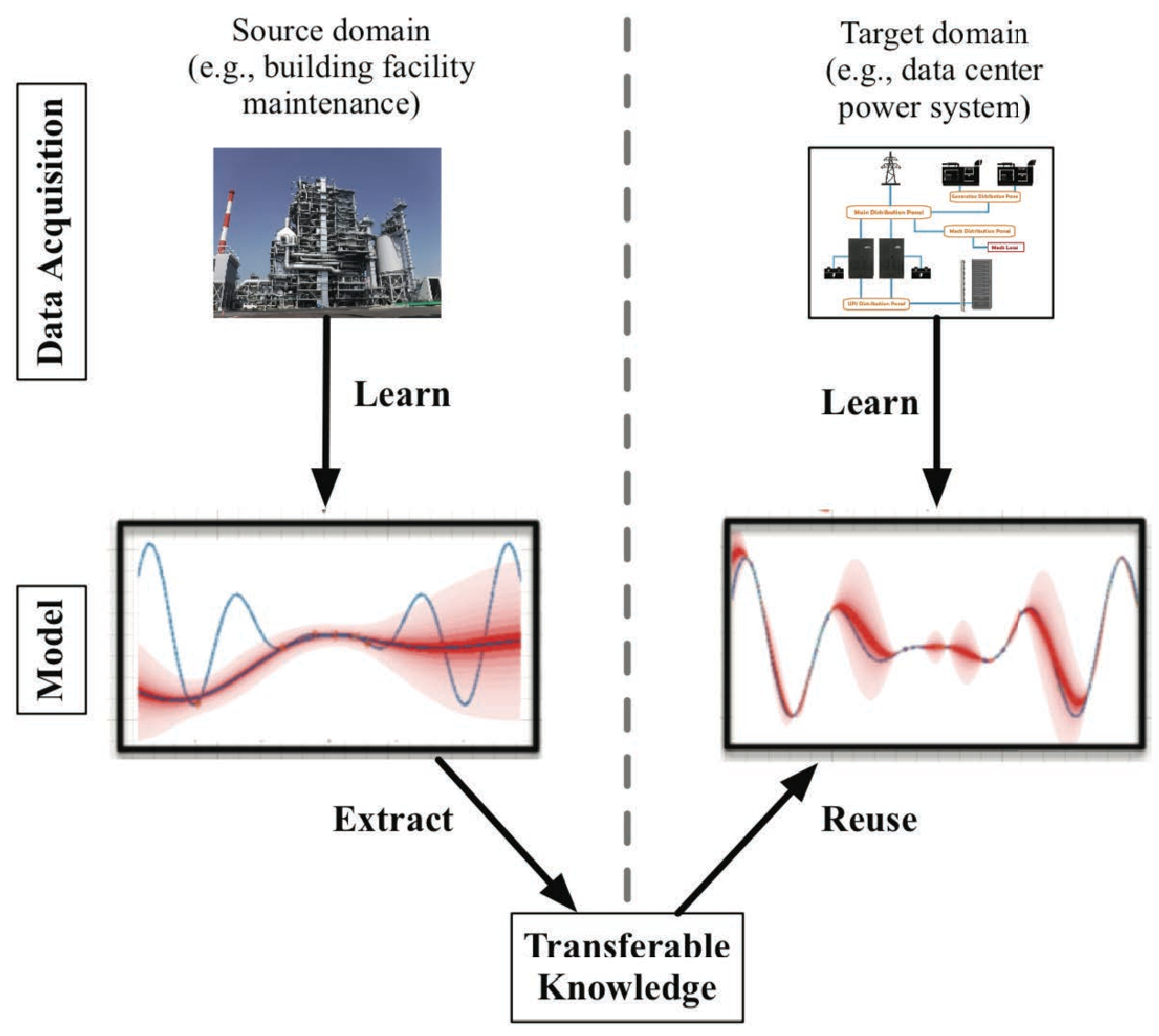}
\caption{The generic framework of transfer learning.}
\label{fig:transfer_arch}
\end{figure}

Transfer learning usually aims to handle the issue of lack of annotated data for the target objects or systems. As we know, the deep learning based approaches (e.g., CNN, RNN, etc.) usually requires a lot of examples of both normal behaviour (of which we often have a lot of) and examples of failures to achieve good performance. However, for a production system, failure events are rare due to the unaffordable and serious consequences when machines running under fault conditions and the potential time-consuming degradation process before desired failure happens. To solve this issue, one method is to use the data augmentation technique -- GAN to generate the training data from a dataset that is indistinguishable from the original data as discussed in the previous subsection. Another method is to employ transfer learning \cite{pan2009survey}. A popular method among all types of transfer learning approaches is domain adaptation which can transfer knowledge from one source domain or a set of source domains to a target domain, as shown in Fig. \ref{fig:transfer_arch}. Whenever the tasks share some fundamental drivers, the transferred knowledge can be used to the target domain and significantly improve its performance (e.g., by reducing the number of samples needed to achieve a nearly optimal performance). Therefore, with labeled data from source domain and unlabeled data from target domain, the distribution discrepancy between the two domains can be mitigated by domain adaptation algorithms.

One of the commonly used domain adaptation methods is representation adaptation which tries to align the distributions of the representations from the source domain and target domain by reducing the distribution discrepancy. In \cite{yang2019intelligent}, Yang \emph{et al.} propose a feature-based transfer neural network (FTNN) to identify faults of bearings used in real-case machines (BRMs) with the help of the knowledge from bearings used in laboratory machines (BLMs). FTNN employs domain-shared CNNs to extract transferable features from raw vibration data of BLMs and BRMs. Then, multi-layer domain adaptation is applied to reduce the distribution discrepancy of the learned transferable features. Finally, pseudo labels are assigned to unlabeled samples in the target domain to train the domain-shared CNN. Guo \cite{guo2018deep} \emph{et al.} propose a deep convolutional transfer learning network (DCTLN) which comprises the condition recognition module and domain adaptation module. The condition recognition module constructs a 1-D CNN to automatically learn features from raw vibration data and recognize health conditions. The domain adaptation module builds a domain classifier and a distribution discrepancy metrics to help learn domain-invariant features. Experimental results show that DCTLN can get the average accuracy of 86.3\%. In \cite{xiao2019domain}, Xiao \emph{et al.} adopt a CNN structure to simultaneously extract the multi-layer features of the raw vibration data from both source domain and target domain. Then, maximum mean discrepancy (MMD) is applied to reduce the distributions discrepancy between two features in the source and target domains. Pang \emph{et al.} \cite{pang2019cross} develop a cross-domain stacked denoising AEs (CD-SDAE) for fault diagnosis of rotating machinery. In this approach, unsupervised adaptation pre-training is utilized to correct marginal distribution mismatch and semi-supervised manifold regularized fine-tuning is adopted to minimize conditional distribution distance between domains.

Parameter transfer, which trains the target network by inheriting parameters from the source network, also has been applied in fault diagnosis. In \cite{he2019improved}, He \emph{et al.} propose deep transfer AE for fault diagnosis of gearbox under variable working conditions with small training samples. An improved deep AE is pre-trained by using sufficient auxiliary data in the source domain, and its parameters are then transferred to the target model. In order to adapt to the characteristics of the testing data, the improved deep transfer is fine-tuned by small training samples in the target domain. Shao \emph{et al.} \cite{shao2018highly} present a CNN-based machine fault diagnosis framework. In this framework, lower-level network parameters are transferred from a previously trained deep architecture, while high-level parameters and the entire architecture are fine-tuned by using task-specific mechanical data. Experimental results illustrate that the proposed method can achieve the test accuracy near 100\% on three mechanical datasets, and in the gearbox dataset, the accuracy can reach 99.64\%. Kim \emph{et al.} \cite{kim2019new} propose a method named selective parameter freezing (SPF) for fault diagnosis of rolling element bearings. Different from the previous approaches, SPF selects and freezes output-sensitive parameters in the layers of the source network, and only allows to retrain unnecessary parameters to the target data. This method provides a new option for the optimization of parameter transfer. In \cite{8957129}, Miao \emph{et. al} provide a novel intelligent method for planetary gearbox fault diagnosis. By dividing the parameters of the classification layer and using a few new fault data to fine-tune the learned network parameters, the proposed method can quickly realize the diagnosis of new type faults while maintaining the original recognition ability.

\begin{figure}[htbp]
\centering
\includegraphics[width=3.5in]{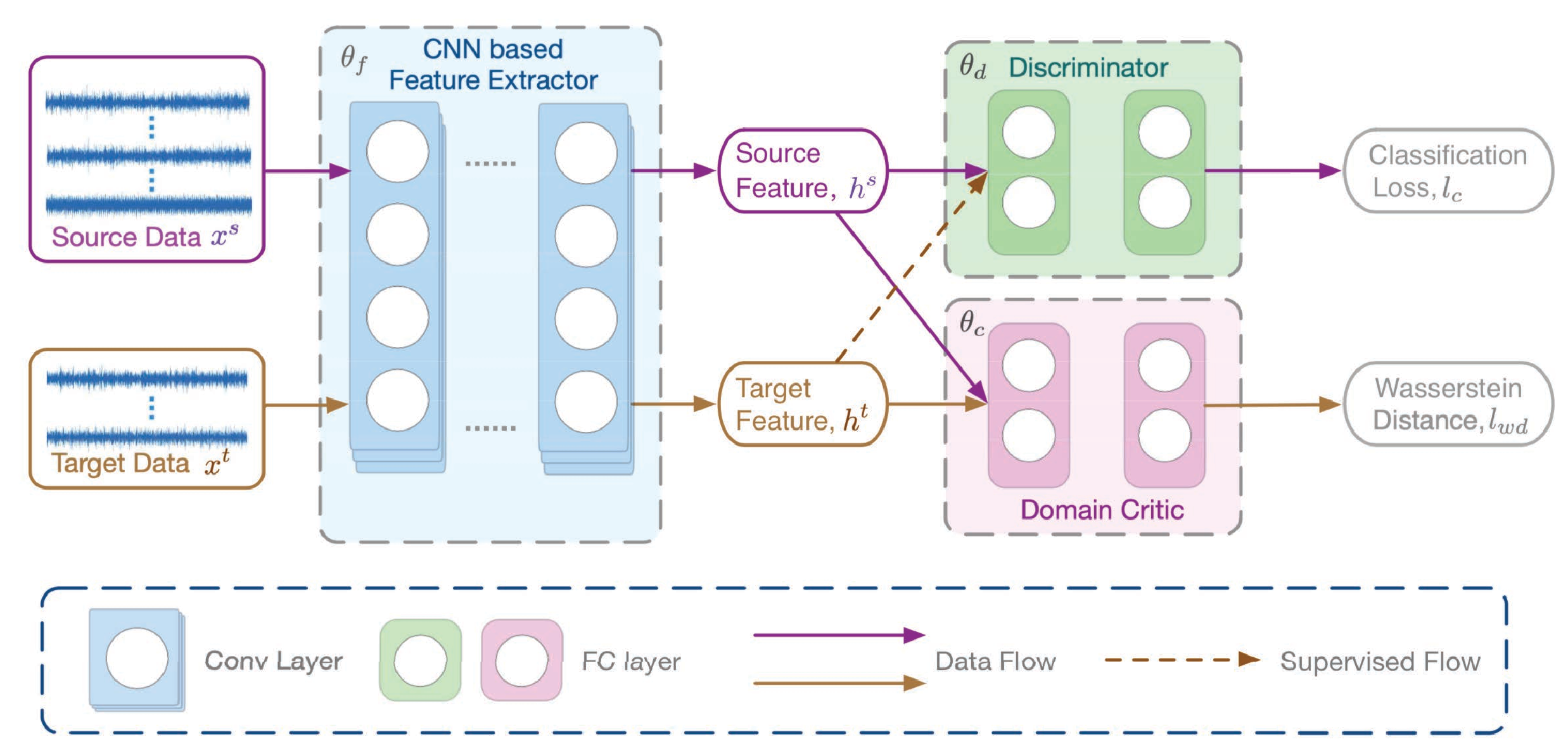}
\caption{The framework WD-DTL for intelligent fault diagnosis \cite{cheng2019wasserstein}. WD-DTL consists of three sub networks: a CNN based feature extractor, a domain critic for learning feature representations via Wasserstein distance, and a discriminator for classification.}
\label{fig:wd-dtl}
\end{figure}

Inspired by GAN, adversarial-based domain adaptation is proposed to minimize the distributions between the source and target domains. In \cite{cheng2019wasserstein}, Cheng \emph{et al.} propose Wasserstein distance based peep transfer learning (WD-DTL) for intelligent fault diagnosis. As shown in Fig. \ref{fig:wd-dtl}, WD-DTL uses source domain labelled dataset to pre-train a CNN model, and then utilizes Wasserstein-1 distance to learn invariant feature representations between source and target domains through adversarial training. Finally, a discriminator with two fully-connected layers is employed to optimize the CNN-based feature extractor parameters by minimizing the estimated empirical Wasserstein distance. Experimental results show that the transfer accuracy of WD-DTL can reach 95.75\% on average. In \cite{lu2019dcgan}, Lu \emph{et al.} develop a domain adaptation combined with deep convolutional generative adversarial network (DA-DCGAN)-based methodology for diagnosing DC arc faults. DA-DCGAN first learns an intelligent normal-to-arcing transformation from the source-domain data. Then by generating dummy arcing data with the learned transformation using the normal data from the target domain and employing domain adaptation, a robust and reliable fault diagnosis scheme based on a lightweight CNN-based classifier can be achieved for the target domain.

\begin{figure}[htbp]
\centering
\includegraphics[width=3.5in]{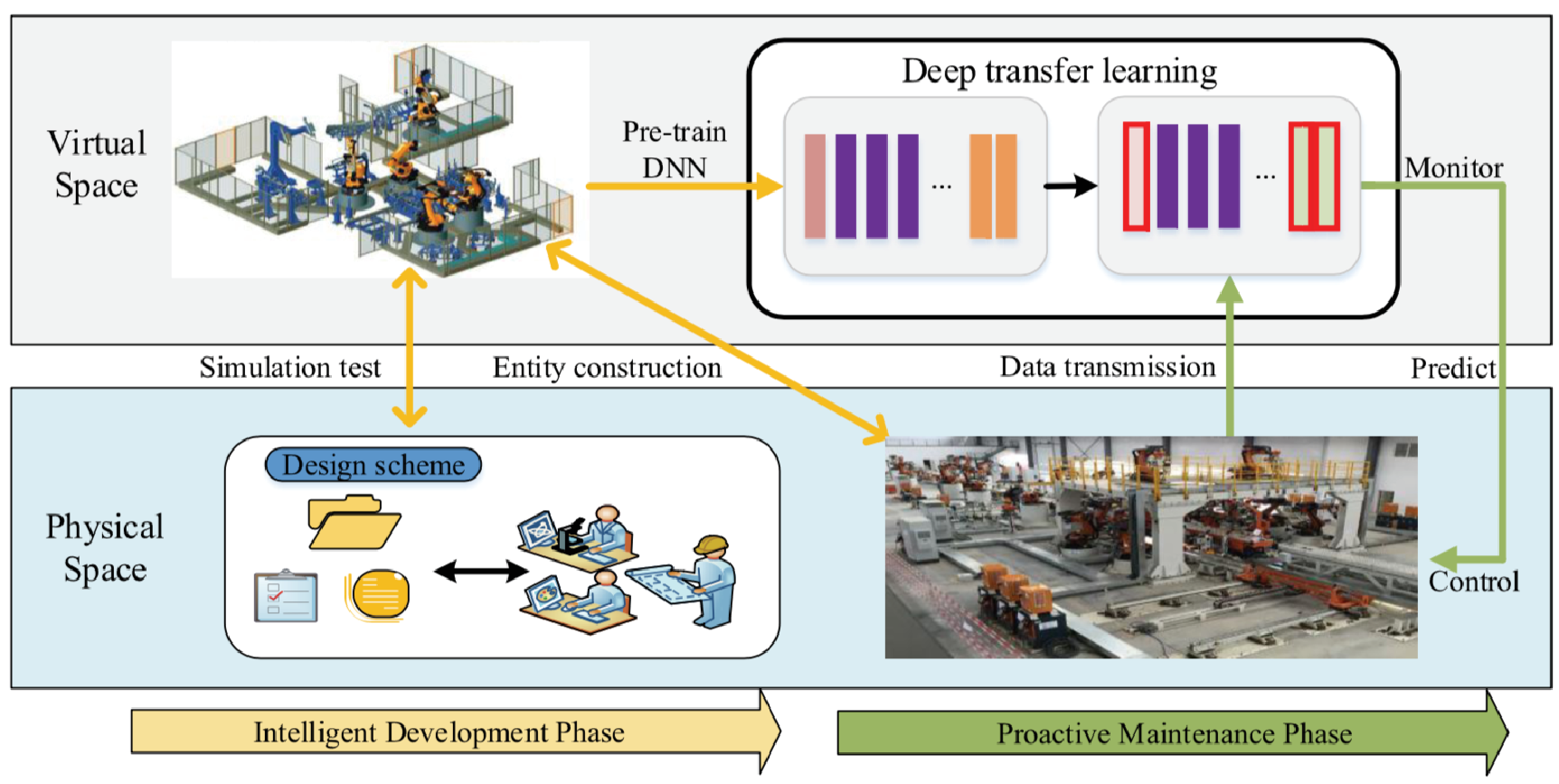}
\caption{Framework of proposed DFDD \cite{xu2019digital}. The yellow and green arrows represent the information flows in two phases, respectively.}
\label{fig:digital_twin_transfer}
\end{figure}

Many other transfer learning based fault diagnosis methods also have been investigated. Xu \emph{et al.} \cite{xu2019digital} present a two-phase digital-twin-assisted fault diagnosis method using deep transfer learning (DFDD), which realizes fault diagnosis both in the development and maintenance phases. As shown in Fig. \ref{fig:digital_twin_transfer}, at first, the potential problems that are not considered at design time can be discovered through front running the ultra-high-fidelity model in the virtual space, while a deep neural network (DNN)-based diagnosis model will be fully trained. In the second phase, the previously trained diagnosis model can be migrated from the virtual space to physical space using deep transfer learning for real-time monitoring and predictive maintenance. This approach ensures the accuracy of the diagnosis as well as avoiding wasting time and knowledge. In \cite{zhang2017new}, Zhang \emph{et al.} present a novel model named WDCNN (Deep CNN with wide first-layer kernels) to address the fault diagnosis problem. The domain adaptation is achieved by feeding the mean and variance of target domain signals to adaptive batch normalization (AdaBN). The domain adaptation experiments were carried out with training in one working condition and testing in another one. WDCNN can get the average accuracy of 90.0\% outperforming FFT-DNN method 78.1\% and the accuracy can be further improved to 95.9\% with AdaBN.

In addition to fault diagnosis, many researchers begin to apply transfer learning for RUL prediction. In \cite{zhang2018transfer}, Zhang \emph{et al.} propose a transfer learning algorithm based on Bi-directional Long Short-Term Memory (BLSTM) recurrent neural networks for RUL estimation. One network was trained on the large amount of data of the source task. Then the learned model was fine-tuned by further training with the small amount of data from the target task, which is usually a different but related task. In the experiments, the task represents the degradation failure under different working conditions. The experimental results show that transfer learning is effective in most cases except when transferring from a dataset of multiple operating conditions to a dataset of a single operating condition, which led to negative transfer learning. Sun \emph{et al.} \cite{sun2018deep} present a deep transfer learning network based on sparse AE (SAE). Three transfer strategies (i.e., weight transfer, transfer learning of hidden feature, and weight update) are employed to transfer an SAE trained by historical failure data to a new object. To evaluate the proposed method, an SAE network is first trained by run-to-failure data with RUL information of a cutting tool in an off-line process. The trained network is then transferred to a new tool under operation for on-line RUL prediction. Similarly, Mao \emph{et al.} \cite{mao2019predicting} propose a two-stage method based on deep feature representation and transfer learning. In the offline stage, a contractive denoising AE (CDAE) is adopted to extract features from marginal spectrum of the raw vibration signal of auxiliary bearings. Then, Pearson correlation coefficient is utilized to divide the whole life of each bearing into a normal state and a fast-degradation state. Finally, a RUL prediction model for the fast-degradation state is trained by applying a least-square SVM. In the online stage, transfer component analysis is introduced to sequentially adapt the features of target bearing from auxiliary bearings, and then the corrected features are employed to predict the RUL of target bearing.

\subsection{Deep Reinforcement Learning (DRL)}
\begin{figure*}[htbp]
\centering
\includegraphics[width=4.5in]{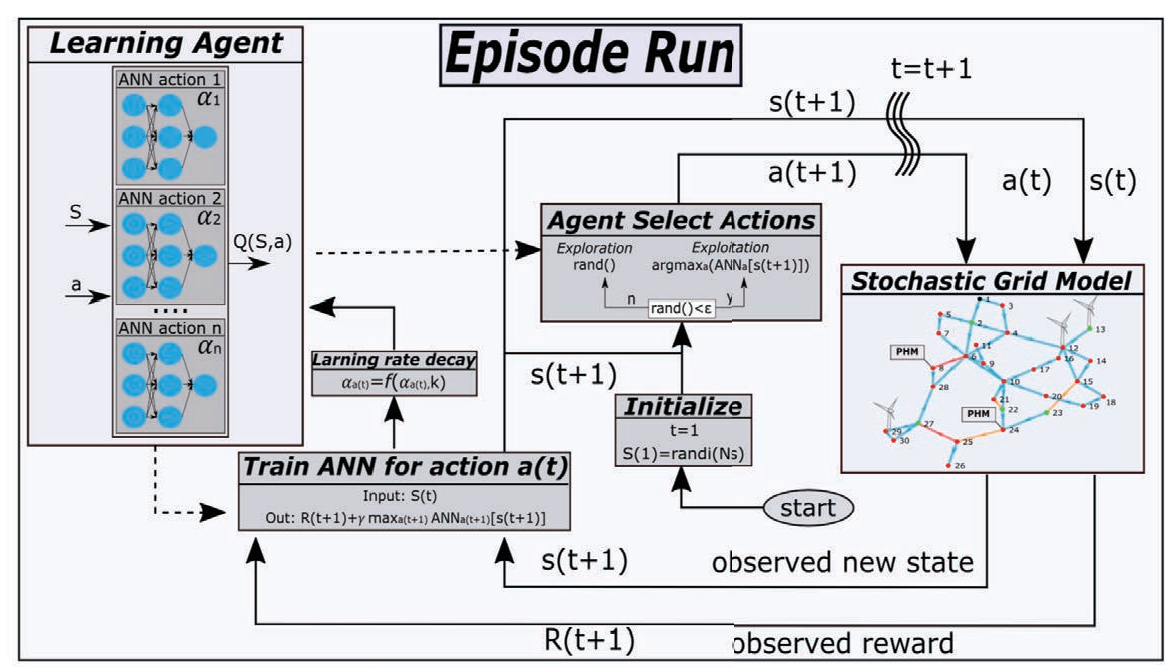}
\caption{The flow chart displays an episode run and how the learning agent interacts with the environment (i.e. the power grid equipped with PHM devices) in the developed RL framework \cite{rocchetta2019reinforcement}.}
\label{fig:drl_arch}
\end{figure*}

Deep reinforcement learning (DRL) is the combination of reinforcement learning (RL) with deep learning and usually used to solve a wide range of complex decision-making tasks. The traditional reinforcement learning algorithm such as Q-learning evaluates the value of the current state $s$ if an action $a$ is taken in this state. The value of a pair $(s,a)$ can be measured by the cost or the profit of taking action $a$ at state $s$. If we can properly determine the value of $(s,a)$ for a sufficiently large number of known state/action pairs, we can choose the optimal control policy by taking the action with the minimum cost or the largest profit. Building a look-up table $Q(s,a)$ to record the value of all known state-action pairs is the key to reinforcement learning approaches. The Q-table is updated by an iterative process during the training. However, Q-learning does not scale well with the complexity of the environment. To address the above scalability issue, the Deep Q-Network (DQN) uses an artificial neural network called Q-network to replace the Q-table (shown in Fig. \ref{fig:drl_arch}). The Q-network, trained to approximate the complete Q-table, can well scale with the number of possible state-action pairs. 

Recently, many research efforts have been devoted to applying DRL to the field of PdM. For example, in \cite{rocchetta2019reinforcement}, Rocchettaa \emph{et al.} develop a DRL framework for the optimal management of the operation and maintenance of power grids equipped with prognostics and health management capabilities. The DRL agent can really exploit the information gathered from prognostic health management devices to select optimal O\&M actions on the system components. The proposed strategy provides accurate solutions comparable to the true optimal. Although inevitable approximation errors have been observed and computational time is an open issue, it provides useful direction for the system operator. To tackle the problem of early classification, Martinez \emph{et al.} \cite{martinez2018deep} propose an original use of reinforcement learning in order to train an end-to-end early classifier agent with simultaneous learning of both features in the time series and decision rules. The experimental results show that the early classifier agent can achieve effective early classification with fast and accurate predictions. In \cite{ding2019intelligent}, Ding \emph{et al.} build an end-to-end fault diagnosis architecture based on DRL that can directly map raw fault data to the corresponding fault modes. Firstly, a stacked AE is trained to sense the fault information existing in vibration signals. Then, a DQN agent is built by sequentially integrating the trained stacked AE and a linear layer to map the output of the stacked AE to Q values. To train the DRL-based algorithm, the authors designed a fault diagnosis simulation environment, which could be considered a ``fault diagnosis game''. Each game contains a certain number of fault diagnosis questions, and each question consists of a fault sample and the corresponding fault label. When the agent is playing the game, the game will have one single question for the agent to diagnose. Then, the game will check whether the agent's answer is correct. If the answer is correct, the reward is increased by $1$, otherwise minus $1$.

Other than end-to-end fault diagnosis and prognosis, DRL also have been employed to derive optimal signal or health indicator (HI). In \cite{dai2020fault}, in order to extract fault characteristics from vibration signal, Dai \emph{et. al} propose a new method which uses DRL algorithm and the reciprocal of smoothness index to control the bandpass filter to select a frequency band with the highest signal-to-noise ratio. Then, envelope demodulation is performed on the filtered signal so as to diagnose the faults of rotating machinery.  Zhang \emph{et al.} \cite{zhang2018equipment} propose a purely data-driven approach for solving the health indicator learning (HIL) problem based on DRL. HIL plays an important role in PdM as it learns a health curve representing the health conditions of equipment over time. The key insight of this paper is that the HIL problem can be mapped to a credit assignment problem. Then DRL learns from failures by naturally back-propagating the credit of failures into intermediate states. In particular, given the observed time series of sensor, operating and event (failure) data, a sequence of health indicators can be derived that represent the underlying health conditions of physical equipment.

\subsection{Typical Hybrid Approaches}
According to the review on DL approaches in the previous subsections, we know that different DL networks have different features and advantages. For example, AE is suitable for high-level feature extraction, LSTM is good at processing sequence data, and DRL can be utilized to learn optimal control policy. Hybrid architectures with multiple types of DL networks usually can achieve a better performance.

\subsubsection{Auto-encoder \& LSTM} 
In \cite{li2019deep2}, Li \emph{et al.} leveraged sparse AE for representation learning and employed LSTM for anomaly identification in mechanical equipment. Experimental results show that the proposed approach could detect anomaly working condition with 99\% accuracy under a completely unsupervised learning environment. In \cite{song2018remaining}, Song \emph{et al.} integrate AE and bidirectional LSTM (BLSTM) to improve the accuracy of RUL prediction for turbofan engines. AE is applied as a feature extractor to compress condition monitoring data, and BLSTM is designed to capture the bidirectional long-range dependencies of features. The RMSE and scoring function values of this hybrid model are 15\% and 23\% lower than those of previous optimal models (MLP, SVR, CNN and LSTM), respectively. 

\subsubsection{CNN \& LSTM} 
In \cite{li2019directed}, Li \emph{et al.} propose a directed acyclic graph (DAG) network that combines LSTM and CNN to predict the RUL of mechanical equipment. The DAG network consists of two paths: LSTM path and CNN path. The output vectors of the two paths will be summed by elements-wise, and then fed into a fully connected layer, which gives the value of the estimated RUL. Pan \emph{et al.} \cite{pan2018improved} combine one-dimensional CNN and LSTM into one unified structure. The output of the CNN is fed into the LSTM for identifying the bearing fault types. The results show that the average accuracy rate in the testing dataset of this proposed method can reaches more than 99\%. In \cite{zhao2017learning}, Zhao \emph{et al.} develop convolutional bidirectional LSTM (CBLSTM) for tool wear prediction. In CBLSTM, CNN is leveraged to extract local features  and bi-directional LSTM is introduced to encode temporal information. Finally, fully-connected layers and the linear regression layer are built on top of bi-directional LSTMs to predict tool wear. Hao \emph{et. al} \cite{hao2020multisensor} present an end-to-end solution with 1D convolutional LSTM networks, where both the spatial and temporal features of multi-sensor measured vibration signals are extracted and then jointed for better bearing fault diagnosis. The experimental results show that the accuracy of the proposed approach can reach 99.86\%-99.99\% with different datasets.

\subsubsection{Auto-encoder \& CNN} 
In \cite{liu2019fault}, Liu \emph{et al.} combine 1-D denoising convolutional AE (DCAE-1D) and 1-D convolutional neural network (AICNN-1D) for the fault diagnosis of rotating machinery. DCAE-1D is utilized for noise reduction of raw vibration signals and AICNN-1D is employed for fault diagnosis by using the output de-noised signals of DCAE-1D. With the denoising of DCAE-1D, the diagnosis accuracies of AICNN-1D can reach 96.65\% and 97.25\% respectively in bearing and gearbox experiments even when $SNR=−2dB$. In \cite{chen2020one}, Chen \emph{et. al} propose one-dimensional convolutional auto-encoder (1D-CAE) for fault detection and diagnosis of multivariate processes. 1D-CAE is employed to learn hierarchical feature representations through noise reduction of high-dimensional process signals. Auto-encoder integrated with convolutional kernels and pooling units allows feature extraction to be particularly effective, which is of great importance for fault detection and diagnosis in multivariate processes.

\subsubsection{Others}
Many other kinds of hybrid approaches are also employed for fault diagnosis and prognosis. For example, in \cite{li2019gear}, the gear pitting fault features are obtained from a 1-D CNN trained with acoustic emission signals and a GRU network trained with vibration signals. Gu \emph{et al.} \cite{yuhai2018research} apply DBN to extract feature vectors of time-series fault data, and leveraged LSTM network to perform fault prediction. In \cite{zhao2019optimal}, Zhao \emph{et al.} construct a novel hybrid deep learning model based on a GRU and a sparse AE to directly and effectively extract features of rolling bearing vibration signals. In \cite{chen2017multisensor}, multiple two-layer sparse AE neural networks are used for feature fusion, a DBN is trained for further classification. In \cite{li2019fault2}, Li \emph{et al.} construct a DBN composed of three pre-trained RBMs to extract features and reduce the dimensionality of raw data. Then, 1D-CNN is applied for further extracting the abstract features and ``softmax'' classifier is employed to identify different faults of rotating machinery.

\subsection{Comparison}
In Section \ref{deep}, many commonly used DL architectures and DL-based approaches have been introduced in literature. In order to provide a quick guidance of how to select an appropriate DL-based method for a specific PdM application, here we briefly describes the advantages, limitations and typical applications of each DL-based approach in Table \ref{tbl:dl} .

\begin{table*}[]
\begin{center}
\caption{Advantages, limitations and typical applications of DL-based Approaches.}
     \begin{tabular}{| p{1.5cm} | p{6cm} | p{4cm} | p{4cm}|}
     \hline
      Networks & Advantages & Limitations & Typical applications
       \\ \hline
	  Auto-encoder
      & 
      \begin{itemize}[leftmargin=*]
      \item No prior data knowledge needed
      \item Can fuse multi-sensory data and compress data
      \item Easy to combine with classification or regression methods
      \end{itemize}
      & 
      \begin{itemize}[leftmargin=*]
      \item Needs a lot of data for pre-training
      \item Cannot determine what information is relevant
      \item Not so efficient in reconstructing compared to GANs 
      \end{itemize}
      &
      \begin{itemize}[leftmargin=*]
      \item Fearture extraction: \cite{jia2018neural, lu2015novel, WANG2018213, zhao2018fault, yuan2017deep}
      \item Multi-sensory data fusion: \cite{chen2017multisensor, ma2018deep}
      \item Fault diagnosis: \cite{shao2017novel, lv2017weighted, mao2019new, haidong2018intelligent, sun2016sparse}
      \item Degradation process estimation: \cite{luo2019early, michau2018feature, michau2018data, wen2018degradation, lin2019novel}
      \item RUL prediction: \cite{xia2018two, ren2018remaining, ma2018predicting, yan2018industrial}
      \end{itemize}
      \\ \hline
      CNN
      &
      \begin{itemize}[leftmargin=*]
      \item Outperforms ANN on many tasks (e.g., image recognition)
      \item Would be less complex and saves memory compared to the ANN
      \item Automatically detects the important features without any human supervision
      \end{itemize}
      &
      \begin{itemize}[leftmargin=*]
      \item Hyperparamter tuning is non-trivial
      \item Easy to overfit
      \item High computational cost
      \item Needs a massive amount of training data
      \end{itemize}
      &
      \begin{itemize}[leftmargin=*]
      \item Fault diagnosis: \cite{ince2016real, kiranyaz2018real, li2019sensor, chen2015gearbox, wang2019deep, oh2019convolutional, jia2019rotating, liu2017infrared}
      \item Degradation process estimation: \cite{guo2018machinery, yoo2018novel, cheng2018online}
      \item RUL prediction: \cite{ren2018prediction, babu2016deep, li2019deep, zhu2018estimation, yang2019remaining, wen2019new}
      \item Joint fault diagnosis and RUL prediction: \cite{liu2019simultaneous}
      \end{itemize}
      \\ \hline
      RNN
      &
      \begin{itemize}[leftmargin=*]
      \item Models time sequential dependencies
      \end{itemize}
      &
      \begin{itemize}[leftmargin=*]
      \item Gradient vanishing and exploding problems
      \item Cannot process very long sequences if using $tanh$ or $relu$ as an activation function
      \end{itemize}
      &
      \begin{itemize}[leftmargin=*]
      \item Fault diagnosis: \cite{li2018intelligent, yuan2019intelligent, zhao2019intelligent, yang2018rotating}
      \item RUL prediction: \cite{chen2019gated, hong2019fault, wu2018remaining, wu2018approach, miao2019joint}
      \item Health indicator construction: \cite{guo2017recurrent, ning2018feature}
      \end{itemize}
      \\ \hline
      DBN
      &
      \begin{itemize}[leftmargin=*]
      \item Has a layer-by-layer procedure for learning the top-down, generative weights
      \item No requirement for labelled data when pre-training
      \item Robustness in classification
      \end{itemize}
      &
      \begin{itemize}[leftmargin=*]
      \item High computational cost
      \end{itemize}
      &
      \begin{itemize}[leftmargin=*]
      \item Fearture extraction: \cite{liang2018bearing, pan2018intelligent, zhang2018analog, shen2019improved}
      \item Fault classification: \cite{tamilselvan2013failure, shao2017rolling, chen2017multisensor, zhu2019novel, wang2018data}
      \item RUL prediction and early fault detection: \cite{deutsch2017using, wang2019early, zhao2017lithium}
      \end{itemize}
      \\ \hline
      GAN
      &
      \begin{itemize}[leftmargin=*]
      \item A good approach to train a classifiers in a semi-supervised way
      \item Does not introduce any deterministic bias compared to auto-encoders
      \item Can be used to address the class imbalance issue
      \end{itemize}
      &
      \begin{itemize}[leftmargin=*]
      \item The training is unstable due to the requirement of a Nash equilibrium
      \item The original GAN is hard to learn to generate discrete data
      \end{itemize}
      &
      \begin{itemize}[leftmargin=*]
      \item Class imbalance issue: \cite{lee2017application, suh2019generative, shao2019generative, wang2019generalization, mao2019imbalanced}
      \item fault identification: \cite{akcay2018ganomaly, jiang2019gan, ding2019generative}
      \item RUL prediction: \cite{khan2018towards}
      \end{itemize}
      \\ \hline
      Transfer Learning
      &
      \begin{itemize}[leftmargin=*]
      \item Saves training time
      \item Does not require a lot of data from the target task
      \item Can learn knowledge from simulations (e.g., digital-twin \cite{xu2019digital})
      \end{itemize}
      &
      \begin{itemize}[leftmargin=*]
      \item Knowledge transfer is only possible when it is 'appropriate' 
      \item Suffers from negative transfer
      \end{itemize}
      &
      \begin{itemize}[leftmargin=*]
      \item  Fault diagnosis: Representation adaptation \cite{yang2019intelligent, guo2018deep, xiao2019domain, pang2019cross}, parameter transfer \cite{he2019improved, shao2018highly, kim2019new}, adversarial-based domain adaptation \cite{cheng2019wasserstein, lu2019dcgan}, digital-twin \cite{xu2019digital}, AdaBN \cite{zhang2017new}
      \item RUL prediction: \cite{zhang2018transfer, sun2018deep, mao2019predicting}
      \end{itemize}
      \\ \hline
      DRL
      &
      \begin{itemize}[leftmargin=*]
      \item Can be used to solve very complex problems
      \item Maintains a balance between exploration and exploitation
      \end{itemize}
      &
      \begin{itemize}[leftmargin=*]
      \item Needs a lot of data and a lot of computation
      \item Assumes the world is Markovian, which it is not
      \item Suffers from the curse of dimensionality
      \item Reward function design is difficult
      \end{itemize}
      &
      \begin{itemize}[leftmargin=*]
      \item Operation and maintenance decision making: \cite{rocchetta2019reinforcement}
      \item Fault diagnosis: \cite{martinez2018deep, ding2019intelligent}
      \item Frequency band selection for Fault Diagnosis: \cite{dai2020fault}
      \item Health indicator learning: \cite{zhang2018equipment}
      \end{itemize}
      \\ \hline
      \end{tabular}
      \label{tbl:dl}
\end{center}
\end{table*}

%% file: future.tex
\section{Future research directions}
\label{future}
In the future, DL techniques will attract more attention in the field of PdM because they are promising to deal with industrial big data. Here the authors list the following research trends and potential future research directions that are critical to promote the application of DL techniques in PdM:
\begin{enumerate}
  \item \textbf{Standards for PdM}: Although there already exist many standards published by different organizations and countries, the emerging technologies have not yet been involved in the standardization under the context of intelligent manufacturing and Industry 4.0. Therefore, it is necessary to draft more standards to normalize the usage of the emerging technologies in PdM, the design of PdM systems, and the workflow for fault diagnosis and prognosis, etc.
  \item \textbf{Large dataset}: The performance of DL-based PdM extremely relies on the scale and quality of the used datasets. However, data collection is time-consuming and costly, it is impractical for some researchers to collect their interested dataset for a specific research target. Therefore, it is meaningful for the PdM community to collect and share large-scale datasets.
  \item \textbf{Data visualization}: As we known, the internal mechanisms of the deep neural networks are unexplainable and it is usually challenging to understand. Therefore, data visualization is essential to analyze massive amounts of fault information in the neural network models and in the learning representations.
  \item \textbf{Class imbalance issue}: For a production system, failure events are rare due to the unaffordable and serious consequences when machines running under fault conditions and the potential time-consuming degradation process before desired failure happens. Therefore, the collected data usually faces the class imbalance issue. Although GAN and transfer learning have been successfully applied to address this issue, it is still challenging to achieve satisfactory performance in many applications and scenarios with imbalanced datasets.
  \item \textbf{Maintenance strategy}: Most of the existing works are devoted to fault diagnosis and prognosis by applying DL techniques, and rarely focus on optimizing maintenance strategy with a certain purpose as described in Section \ref{obj}. However, it is significant to properly schedule the maintenance activities by applying AI technologies (e.g., DRL) for maintenance automation, cost saving as well as downtime reduction.
  \item \textbf{Hybrid network architecture}: According to the comprehensive review on DL approaches in this paper, we know that different DL networks have different features and advantages. For example, auto-encoder is suitable for high-level feature extraction, LSTM is good at processing sequence data, and DRL can be used to learn optimal control policy. Therefore, more hybrid network architectures can be explored and designed to achieve remarkable performance in some complicated applications (e.g., fault diagnosis for multi-component systems).
  \item \textbf{Digital twin for PdM}: A digital twin usually comprises a simulation model which will be continuously updated to mirror the states of their real-life twin. This new paradigm enables us to obtain extensive data regarding run-to-failure data of critical and relevant components. This will be highly beneficial and necessary for successful implementations of fault detection and prediction. For example, a two-phase digital-twin-assisted fault diagnosis method using deep transfer learning is proposed in \cite{xu2019digital}. 
  \item \textbf{PdM for multi-component systems}: With the fast growth of economy and the development of advanced technologies, manufacturing systems are becoming more and more complex, which usually involve a high number of components. However, most of the existing DL-based approaches only focus on the fault diagnosis and prognosis for a specific component. Multiple components and their dependencies will increase the complexity and difficulty of a DL-based PdM algorithm. Therefore, how to design an effective  DL-based PdM algorithm for multi-component systems is still an open issue. 
\end{enumerate}

%% file: conclusion.tex
\section{Conclusion}
\label{conclusion}
This paper presented a comprehensive survey of PdM system architectures, maintenance purposes and learning-based approaches. First, we presented an overview of the PdM system architectures. This provided a good foundation for researchers and practitioners who are interested to gain an insight into the PdM technologies and protocols to understand the overall architecture and role of the different components and protocols that constitute the PdM systems. Then, we introduced three types of purposes for performing PdM activities including cost minimization, availability/reliability maximization and multiple objectives. Afterwards, we provided a review of the existing learning-based approaches that comprise: traditional ML-based and DL-based approaches. Special emphasis is placed on the DL-based approaches that have spurred the interests of academia for the past five years. Finally, we outlined some important future research directions that are critical to promote the application of DL techniques in the context of PdM.